\documentclass[fleqn,usenatbib]{mnras}

\usepackage[T1]{fontenc}
\usepackage{ upgreek }
\DeclareRobustCommand{\VAN}[3]{#2}
\let\VANthebibliography\thebibliography
\def\thebibliography{\DeclareRobustCommand{\VAN}[3]{##3}\VANthebibliography}

\usepackage{graphicx}	% Including figure files
\usepackage{amsmath}	% Advanced maths commands
\usepackage{amssymb}	% Extra maths symbols
\usepackage{xspace}
\usepackage{enumitem}

\usepackage{scalerel,stackengine,amsmath}

\DeclareMathOperator*{\argmin}{arg\,min}

\usepackage{orcidlink}
\usepackage{algorithm}
\usepackage{dsfont}
\usepackage{soul}  % for strikeout
\newcommand{\crit}{{\rm crit}}
\newcommand{\eg}{\textit{e.g.}}
\newcommand{\ie}{\textit{i.e.}}

\usepackage[noend]{algpseudocode}
\usepackage{newtxtext,newtxmath}
\makeatletter
\def\algbackskip{\hskip-\ALG@thistlm}
\makeatother

\title[Optimised Samples for WL Cluster Masses]{Reducing Model Error Using Optimised Galaxy Selection: Weak Lensing Cluster Mass Estimation  }

\author[]{Markus~Michael~Rau$^{1,2}$\thanks{E-mail: markus.rau@newcastle.ac.uk}\orcidlink{0000-0003-3709-1324}, Florian~K\'eruzor\'e$^{2}$\orcidlink{0000-0002-9605-5588}, Nesar~Ramachandra$^{3}$,   Lindsey~Bleem$^{2}$ \orcidlink{0000-0001-7665-5079} 
\\
$^{1}$School of Mathematics, Statistics and Physics,Newcastle University, Newcastle upon Tyne, NE17RU, United Kingdom\\
$^{2}$High Energy Physics Division, Argonne National Laboratory, Lemont, IL 60439, USA\\
$^{3}$Computational Science Division, Argonne National Laboratory, Lemont, IL 60439, USA\\
}

\date{Accepted XXX. Received YYY; in original form ZZZ}

\pubyear{2024}

\begin{document}
\label{firstpage}
\pagerange{\pageref{firstpage}--\pageref{lastpage}}
\maketitle

\begin{abstract}
Galaxy clusters are one of the most powerful probes to study extensions of General Relativity and the Standard Cosmological Model. Upcoming surveys like the Vera Rubin Observatory's Legacy Survey of Space and Time are expected to revolutionise the field, by enabling the analysis of cluster samples of unprecedented size and quality. To reach this era of high-precision cluster cosmology, the mitigation of sources of systematic error is crucial. A particularly important challenge is bias in cluster mass measurements induced by the mismodelling of photometric redshift estimates of source galaxies. This work proposes a method to optimise the source sample selection in cluster weak lensing analyses drawn from wide-field survey lensing catalogs to reduce the bias on reconstructed cluster masses. We use a combinatorial optimisation scheme and methods from variational inference to select galaxies in latent space to produce a probabilistic galaxy source sample catalog for highly accurate cluster mass estimation. We show that our method reduces the critical surface mass density $\Sigma_\crit$ relative modelling bias on the 60-70\% level, while maintaining up to 90\% of galaxies. We highlight that our methodology has applications beyond cluster mass estimation as an approach to jointly combine galaxy selection and model inference under sources of systematics.
\end{abstract}

\begin{keywords}
cosmology: observations -- galaxies: distances and redshifts -- methods: data analysis -- gravitational lensing: weak -- clusters: general
\end{keywords}

%%%%%%%%%%%%%%%%%%%%%%%%%%%%%%%%%%%%%%%%%%%%%%%%%%

%%%%%%%%%%%%%%%%% BODY OF PAPER %%%%%%%%%%%%%%%%%%
\section{Introduction}
\label{sec:1}

%%% Cluster cosmology context
The abundance of massive dark matter halos in mass and redshift is known to be a powerful cosmological probe.
The most massive halos are located at the intersection of cosmic filaments and host galaxy clusters, which can be detected in sky surveys and used to set constraints on cosmological parameters \citep[see, \eg,][for reviews]{2005RvMP...77..207V,2011ARA&A..49..409A}.
In particular, cluster abundances have been shown to be a competitive probe of matter distribution properties on cosmic scales (through cosmological parameters $\Omega_{\rm m}$ and $\sigma_8$), of the sum of neutrino masses, and of dark energy properties \citep[\eg,][]{planck15_clcosmo, desy1_clcosmo, kidsdr3_clcosmo, salvati22, bocquet24ii,Ghirardini24}, as well as an efficient test of modified gravity models \citep[\eg,][]{2018IJMPD..2748006C}.

%%% Cluster detection
Broadly speaking, cluster-based cosmological analyses require three steps.
First, a large sample of galaxy clusters needs to be detected from survey data.
The main ways to achieve this goal are the detection of clusters through their imprint on the cosmic microwave background \citep[through the Sunyaev-Zeldovich effect \citep{sunyaev72}; see][for a review]{2019SSRv..215...17M}; the identification of spatial overdensities in the galaxy distribution field \citep[\eg,][]{redmapper,amico}; and/or the detection of extended X-ray emission from the intracluster plasma \citep[see, \eg,][for a review]{2013AN....334..482B}.
These methods have been used to detect clusters in a variety of cosmological surveys, resulting in samples of hundreds to thousands of clusters \citep[\eg,][]{desy1clcat, hilton21, kidsdr3_clcat, bleem24, bulbul24}.
In the near future, several tens of thousands of clusters are expected to be detected in upcoming cosmological surveys such as the Vera Rubin Observatory's Legacy Survey of Space and Time \citep[LSST,][]{2009arXiv0912.0201L}, \textit{Euclid} \citep{2019A&A...627A..23E}, SPT-3G \citep{2014SPIE.9153E..1PB}, Simons Observatory \citep{2019JCAP...02..056A}, and CMB-S4 \citep{2016arXiv161002743A}.
%LB I added the published eROSITA catalog so cut ref eROSITA \citep{2012MNRAS.422...44P},

%%% Cluster mass calibration
The next major steps in the cosmological exploitation of cluster samples entail characterizing the selection function \citep[see \eg,][]{costanzi19,bleem24,clerc24} and calibrating the masses of the selected clusters \citep[see, \eg,][for a review]{2019SSRv..215...25P}. These latter steps are crucial, since they enable connections to the theoretical predictions that allow for cosmological parameter inference. 
One of the most widely used methods to estimate cluster masses is via measurements of weak gravitational lensing in optical datasets \citep[see][for a review]{umetsu20}, where measuring the distortion of galaxies behind clusters enables the reconstruction of their gravitational potential. This can then be used to estimate cluster masses; either on a per-cluster basis or on average in bins of redshift and survey observable.

%%% Systematics in WL mass measurements
While being one of the most accurate methods to measure the mass of galaxy clusters (in particular using photometric large sky surveys), weak lensing cluster mass calibration is subject to several sources of systematic error, including, \eg, miscentering of deprojected cluster shear profiles, contamination of the lensing signal by cluster members, covariances between lensing signal and survey observables, intrinsic alignment, and halo mass modelling \citep[\eg,][]{grandis21, sommer22, bocquet23i}.
Another major source of systematic uncertainty is the calibration of photometric redshifts for lensed galaxies.
Recently, \citet{bocquet23i} showed that for a sample of clusters selected in surveys from the South Pole Telescope with weak lensing mass calibration from the Dark Energy Survey, uncertainties on source photometric redshifts were the leading source of uncertainty on the bias of the weak lensing mass estimator for cluster redshifts $z > 0.45$ \citep[see fig. 10 of][]{bocquet23i}.
While the cosmological constraints derived from this sample are dominated by statistical uncertainties and shape noise in lensing measurements \citep{bocquet24ii}, these will be reduced in upcoming cosmological datasets.
Ensuring the accuracy of photometric redshift estimates is therefore crucial for the future of cluster cosmology.

Estimators of redshifts that use the photometry of galaxies \citep[for a recent review, see][]{2019NatAs...3..212S, 2022ARA&A..60..363N} can be categorized into empirical, or machine learning (ML) methods \citep{2003LNCS.2859..226T, 2004PASP..116..345C, 2010ApJ...715..823G, 2013MNRAS.432.1483C,  2015MNRAS.449.1043B, 2015MNRAS.452.3710R, 2016A&C....16...34H} that use training data from spatially overlapping regions between photometric and spectroscopic surveys and template fitting methods \citep[\eg,][]{1999MNRAS.310..540A, 2000ApJ...536..571B,2006A&A...457..841I, 2006MNRAS.372..565F, 2015MNRAS.451.1848G, 2016MNRAS.460.4258L, 2020arXiv200712178M} that fit spectral energy distribution (SED) models to the observed photometry. Both approaches are complementary but fundamentally limited by either the availability of complete training data or epistemic uncertainty in the SED and selection function models \citep[see \eg][]{10.1093/mnras/stu1424, 2015APh....63...81N, 2017ApJ...841..111M, 2019ApJ...877...81M, 2020MNRAS.496.4769H}. As such, despite  significant work, photometric redshift estimation remains a challenge for future surveys owing to a number of factors summarized in the following. 

Large area photometric surveys like LSST extract redshift information using galaxy images in a set of broad optical filter bands, whose response are defined by the filter transmission functions. The galaxy flux (which is an integral of the product between filter transmission function and SED) and its uncertainty depend on the data reduction and observing strategies. While the shape and magnitude of the SED depends strongly on the wavelength, redshift and on a number of latent variables (for instance, the “hidden” variables related to the physical properties of the galaxies and their stellar components), the small number of broad band filters at fixed wavelength ranges lead to a significant information loss. As a result, small changes in the measured broad band flux can map to a large number of possible solutions in the aforementioned latent parameter space. 
This is further complicated by epistemic uncertainty in e.g. stellar-population-synthesis models \citep[SPS; see \eg,][for a review]{conroy13} or selection function models, particularly in regions of colour-magnitude space where high-resolution spectroscopic calibration data is costly to obtain. 

These challenges result in photometric redshift estimates with substantial and complex uncertainty. Figure~1 in \citet{10.1093/mnras/stz2162} is an excellent illustration of this problem, where two galaxy SEDs that differ in galaxy type and redshift would yield similar optical flux measurements. 

Thus, broad band optical surveys cannot constrain the shape of galaxy spectral energy distributions to very high precision and therefore rely on spectroscopic surveys or select fields with highly accurate, narrow filter, multiband photometry like COSMOS \citep{2020arXiv200711132A} to calibrate spectral energy distribution models or train empirical mappings from broad band photometry to redshift. If these samples are incomplete, they can produce a bias in the derived redshift \citep{2015APh....63...81N, 2020MNRAS.496.4769H} that cannot be easily resolved by adding additional parameters because, as mentioned above, the constraining power of broad band photometry is limited. We will refer to these biases as misspecification error.   

It has become clear in recent years that the estimation of redshifts using broad-band photometry is not sufficient to meet the high fidelity required by future large-area photometric surveys like LSST \citep{2018arXiv180901669T}. As a result, the inclusion of redshift information derived from spatial cross-correlations between the photometric and spectroscopic surveys has become a highly relevant research topic \citep[\eg,][]{2008ApJ...684...88N, 2013arXiv1303.4722M, 2013MNRAS.433.2857M, 2016MNRAS.462.1683S, 10.1093/mnras/stx691, Morrison2016, 2017arXiv171002517D, 2018MNRAS.477.1664G, 2020A&A...642A.200V, 2020MNRAS.491.4768R, 2021A&A...647A.124H}. This cross-correlation signal measured between the photometric sample and spectroscopic galaxy sample selected to lie within the redshift bin $\Delta z$ is roughly speaking a product of the fraction of galaxies in the photometric sample that lies within $\Delta z$ and the redshift-dependent galaxy-dark matter bias of the photometric sample. The degeneracy between these two modelling components is an important challenge in cross-correlation redshift estimation. Thus, combining the redshift information derived from photometric observations and spatial clustering measurements is nowadays standard in all current Stage~III \citep{detf06} surveys like the Dark Energy Survey \citep{10.1093/mnras/stz2162, 2021MNRAS.505.4249M,2022MNRAS.513.5517C, 2022MNRAS.510.1223G, 2024MNRAS.527.2010G}, the Hyper Suprime Cam Survey \citep{2023MNRAS.524.5109R} and the Kilo Degree Survey \citep{2021A&A...647A.124H}. 

While it is crucial to overcome these challenges in redshift estimation to achieve the full constraining power of next-generation surveys, there are immediate opportunities to improve constraints on some of the current-generation survey data via meticulous construction of photometric galaxy samples optimised for the probe of interest.  This is especially true for probes with a high signal-to-noise data vector, like cluster-galaxy weak lensing (which has a relatively high significance compared to, \eg, galaxy-galaxy lensing). 
When estimated using wide-field survey data, the lensing samples adopted for cluster mass calibration have been typically drawn from catalogs optimised for cosmic shear and galaxy-galaxy lensing analyses \citep[\eg][]{bocquet23i}. In such cases, refining the adopted lensing sample could reduce systematic biases from faulty photometric redshift estimates, even if it leads to a reduction in sample size.
%
%
%Given the intrinsic challenges in photometric redshift estimation and the high signal-to-noise in cluster mass estimates, makes a reduction in the effective sample size of background galaxies a viable option if systematic bias from faulty photometric redshift estimates is reduced. This is especially true for probes with a high signal-to-noise data vector like weak lensing cluster mass estimates.
In this work, we propose a combinatorial optimisation scheme for source sample selection to minimize the bias on surface density estimates for cluster-scale halos, while preserving a large background galaxy sample to maximize the statistical power of weak lensing mass calibration.
This work is based on the premise that while the aforementioned challenges in photometric redshift calibration can arguably not provide a consistent estimate of redshift bias, it can provide a localization of model misspecification in parameter space as well as a relative measure of model accuracy via e.g. comparing several different photometric redshift codes. We therefore do not assume that the photometric redshift error is quantifiable and thus our main optimization objective is not a traditional bias-variance trade-off. We instead focus on utilizing metrics that provide an assessment of risk for biases in quantities of interest that are offset by the galaxy sample size. Since redshift calibration is strongly survey dependent, we will conduct a mock study with known ground truth. However, we stress that in the empirical setting we will not have access to this ground truth and all statements made henceforth quantifying a `bias' (using conventions in the field of cosmology), in quantities of interest or sample redshift distributions are to be understood within the aforementioned limitations of redshift calibration. 

This article is structured as follows: we quantify the impact of photometric redshift error on cluster mass measurements (\S\ref{sec:3}), discuss our photometric redshift error model (\S\ref{sec:2}), introduce our statistical inference methodology (\S\ref{sec:unfolding_sample_selection}), and present our approach for sample selection optimisation (\S\ref{sec:4}). \S\ref{sec:5} presents and discusses the results of our mock analysis. \S\ref{sec:discussion} provides a discussion on how the presented methodology can be used in an upcoming cluster weak lensing mass measurement analysis. \S\ref{sec:illustrative_mock} illustrates how the output from our method can be used in cluster mass estimation using a mock study. \S\ref{sec:6} then closes the paper with a summary and discussion of future work. 

\textit{Conventions:} we note that, where applicable, we assume a fiducial $\Lambda$CDM cosmology with $\sigma_8=0.8$, $\Omega_b = 0.05$, $\Omega_c = 0.25$, $h=0.7$, and $n_s(k_s=0.002) = 0.95$. Magnitudes are reported in the AB system \citep{oke74}. Observed quantities, i.e. data, are denoted using a `hat'. We note that realizations of random variables (denoted in captial letters) are written in lower case. The variables in bold denote vector quantities.

\begin{table}
\centering
\begin{tabular}{ c c c c } 
\hline\hline
\multicolumn{3}{|c|}{\textbf{Variables}}\\\hline\hline
Notation & Definition & Description \\\hline
$(\mathbf{\vec{Z}_{p}}/\mathbf{\vec{Z}})$ & $\mathbb{R}^{N_{\rm samp}}$ & (Photometric/True) Redshift \\
($N_{\rm bins}/N_{\rm gal}$) & $\mathbb{N}$ & Number of (histogram bins/galaxies) \\
$\sigma$ & $\mathbb{R}^{+}$ & Photometric redshift error (Tab.~\ref{tab:2}) \\
$z_{\rm bias}$ & $\mathbb{R}$ & Photometric redshift bias (Tab.~\ref{tab:2}) \\
$\mathbf{\phi}$ & $\mathbb{S}^{N_{\rm bins}}$ & Sample-pz parameters (Eq.~\ref{eq:sample_pz}) \\
$\boldsymbol{\alpha}$ & $\mathbb{R}^{+}$ & Concentration parameter  \\
$\mathcal{D}$ & $\mathbb{R}^{N_{\rm samp}}$ & Observed Data \\
$\Sigma_{\rm crit}$ & $\mathbb{R}^{+}$ & Critical surface mass density \\
$\Sigma_{\rm crit, (fid/bias)}$ & $\mathbb{R}^{+}$ & $\Sigma_{\rm crit}$ for (fiducial/biased) parameters Tab.~\ref{tab:2} \\\hline\hline
\multicolumn{3}{|c|}{\textbf{Distributions}}\\\hline\hline
Notation & Definition & Description \\\hline
$q(\mathbf{\phi} | \boldsymbol{\alpha})$ & $\mathbb{S}^{N_{\rm bins}} \rightarrow \mathbb{S}^{1}$ & Dirichlet variational distribution \\
$Dir(\mathbf{\phi} | \boldsymbol{\alpha})$ & $\mathbb{S}^{N_{\rm bins}} \rightarrow \mathbb{S}^{1}$ & Dirichlet distribution \\
$\mathcal{U}(z_L, z_R)$ & $\mathbb{R}^{+} \rightarrow \mathbb{S}^{1}$ & Uniform distribution in $(z_L, z_R)$ \\
\hline
\end{tabular}
\caption{\label{tab:1}  Glossary of variable names and definitions used in this work, presented in separate sub-tables for variables and distributions. The first column lists the variable name. The second lists the set in which the variable is an element or the domain/codomain of functions as defined in this work. The last column provides a short description of the variable. $\mathbb{S}^{d}$ denotes the $d$-dimensional probability simplex and $\mathbb{N}$ the set of natural numbers.  }  
\end{table}

\section{Quantifying Cluster Mass Measurement systematics induced by Photo-Z error}
\label{sec:3}
Weak lensing mass measurements are one of the most reliable methods to perform the mass calibration of large cluster samples for cosmology. Consider a cluster that acts as a lens for the light emitted by source galaxies, that are located behind the cluster. The gravitational lensing effect distorts the shape of source galaxies, creating a measured tangential shear $\gamma_{\rm tan}(R)$ within a projected radius $R$ from the cluster center proportional to the cluster mass, or more precisely, the excess mass density 
$\left\langle \Sigma(r) \right\rangle_{r < R} - \overline{\Sigma}(R)$: 
\begin{equation}
	\left\langle \Sigma(r) \right\rangle_{r < R} - \overline{\Sigma}(R) = \Sigma_\crit \, \gamma_{\rm tan}(R) \, .
 \label{eq:cluster_mass_estim}
\end{equation}
The factor of proportionality is the critical surface density $\Sigma_\crit$ that depends on the geometry of the source-lens system and therefore requires accurate modelling of the source sample redshift distribution (sample-pz). In this work we will use the critical surface density averaged over the sample-pz defined as:
\begin{equation}
\Sigma_\crit  \propto \int_{z_{\rm Lens}}^{\infty} {\rm d}z \, p(z) \left( \frac{D_{\rm s}(z)}{D_{\rm d}(z_{\rm Lens}) D_{\rm ds}(z_{\rm Lens}, z)} \right) \, .
\label{eq:sig_crit}
\end{equation}
Here $z_{\rm Lens}$ denotes the redshift of the lens or cluster and the lens-source geometry is described by $D_{\rm d}$, $D_{\rm s}$ and $D_{\rm ds}$, which denote the angular diameter distances to the lens, the source and between the lens and the source. We note that Eq.~\ref{eq:sig_crit} can be interpreted as the expected critical surface density with respect to a sample redshift distribution $p(z)$ behind the lens at redshift $z_{\rm lens}$. We refer to \citet[][p.~48]{2001PhR...340..291B} for more details. We see that the integrand of Eq.~\ref{eq:sig_crit} depends on the source sample redshift distribution. In the following, we assume that the sample-pz is estimated using photometric data. We note that in real observational studies this truncated expectation value is evaluated with respect to the inverse $\Sigma_\crit$ \citep[see.][Eq. 95]{umetsu20}. We have confirmed that both choices lead to very similar results in \S~\ref{sec:5}.
%\lbnotes{TBD: add citation to Bartlemann somewhere in these equations}

%\lbnotes{Nothing in the equations alone imply a 5-10\% error, do you mean to say "Given the typical biases in photometric redshifts from BLAH, following these 2 equations this would translate into a X\% mass bias.}
As implied by Eq.~\ref{eq:cluster_mass_estim} and Eq.~\ref{eq:sig_crit}, systematic errors in 
the estimation of the sample-pz propagate into the modelling of the cluster mass. Given the photometric redshift error in the recent analysis \citep[see \eg,][]{2014MNRAS.439...48A}, we can expect this error to make up around $5\%$ of the uncertainty on derived cluster masses, which makes it one of the dominant sources of systematic error in weak lensing cluster mass estimation \citep[see fig. 10 of][]{bocquet23i}.

\section{Data Model: Photometric Redshifts }
\label{sec:2}

In the introduction, photometric redshift estimation was discussed as a complex inference problem that requires the combination of information from the clustering of galaxies and photometry. The inference is challenging due to the necessary high accuracy in SPS modelling, the modelling of selection functions, and astrophysical systematics like galaxy-dark matter bias. The goal of this section is not to accurately reproduce this process, but rather to produce a biased $p(z)$ with sufficient realism to demonstrate the value of our proposed sample selection methodology. Tab.~\ref{tab:1} provides a glossary of variable names and definitions used in the following. 

% \lbnotes{in order to produce a p(z) [word choice..] realisation of sufficient realism to demonstrate the value of our proposed optimisation technique.}
%using a simplified model. Agreed

For this purpose, we use a heteroscedastic additive error model (see Tab.~\ref{tab:1}) for photometric redshift where $(Z_{\rm p, i}/Z_{\rm i})$ describes the (photometric/true) redshift of galaxy $i$ in a sample of $N_{\rm samp}$ galaxies\footnote{We will denote random variables using capital letters and their realizations in lower case.}:

\begin{align}
    Z_{\rm p, i} &= Z_{\rm i} + \epsilon_{\rm i} \\
    \epsilon_{\rm i} &\sim  \mathcal{N}(z_{\rm bias,i}, \sigma_i) \, . 
\end{align}

Our goal is to infer the population distribution $p(z)$. In the following we consider the values of the random variables $Z_{\rm p, i}$ as observed, of $\epsilon_{\rm i}$ as estimated, and of $Z_{\rm i}$ as latent.  The additive noise $\epsilon_{i}$ for each galaxy $i$ is assumed to be a normal distribution with mean $z_{\rm bias, i}$ and standard deviation $\sigma_{i}$. 
\begin{table}
\centering
\begin{tabular}{c c c } 
\hline\hline
  &  \multicolumn{2}{|c|}{\textbf{Redshift distribution}}\\\hline\hline
  & Fiducial & Biased \\ [0.5ex] 
 \hline
 $z_{\rm bias}(z)$ & $0$ &  $-0.106 \cdot z + 0.206 \cdot z^2$  \\ 
 $\sigma(z)$ & $0.02 \cdot (1 + z)$ &  $0.02 \cdot (1 + z)$ \\ 
 \hline
\end{tabular}
\caption{\label{tab:2} Fiducial and biased parameter values considered in this work. The columns list the parameter names and the specification of fiducial and biased parameters. The rows list the parameters considered in this work, where $z_{\rm bias}(z)$ and $\sigma(z)$ are functions of the true galaxy redshifts $z$. The difference between the fiducial and biased parameter set lies in the different prescriptions of $z_{\rm bias}(z)$. }
\end{table}

Tab.~\ref{tab:2} summarizes the parameter values considered in this work, where we study a (``fiducial''/``biased'') set of parameter values. These parameter values have been selected to mimic photometric redshift errors studied in \citet{2020MNRAS.496.4769H} for a severe but realistic case of model misspecification induced by incomplete spectroscopic redshift training data (cf.~\S~\ref{sec:1}). 
%The effect of the biased parameter values on the results will be considered in relation to the results of the fiducial parameters. 
%\lbnotes{Where did the biased model come from? Need citation and motivation.}

The individual galaxy likelihood is then modeled as: 

\begin{equation}
    p(\hat{z}_{\rm p, i} | z_i) = \mathcal{N}\left(\hat{z}_{\rm p, i} | \mu=z_i+z_{\rm bias}(z_i), \sigma=\sigma(z_i)\right) \, .
\end{equation}

Assuming that the individual galaxy redshifts are independent, we write the joint likelihood as:
\begin{equation}
    p(\boldsymbol{\hat{z}_{\rm p}} | \boldsymbol{z}, \boldsymbol{\sigma}, \boldsymbol{z_{\rm bias}}) = \prod_{i = 1}^{N_{\rm samp}} p(\hat{z}_{\rm p, i} | z_i, \sigma_i, z_{\rm bias, i})  \, , 
      \label{eq:joint_indiv_like}
\end{equation}
where ($\hat{z}_{\rm p, i}$/$z_i$) denotes the (observed/latent) realisations of ($\mathbf{Z_{\rm p, i}}$/$\mathbf{Z_{\rm i}}$).

We parametrize the sample redshift distribution $p(z| \boldsymbol{\phi_{\rm nz}})$ as a histogram with $N_{\rm bins}$ normalized bin heights given by the parameter vector $\boldsymbol{\phi_{\rm nz}}$ 
\begin{equation}
    p(z | \boldsymbol{\phi_{\rm nz}}) = \sum_{i = 1}^{N_{\rm bins}} \phi_\text{nz, i} \mathcal{U}_{i}(z) \,
    \label{eq:sample_pz}
\end{equation}
where $\mathcal{U}_{i}(z)$ is the uniform distribution between the (left/right) edges ($z_{\rm i, L}$/$z_{\rm i, R}$) of redshift bin $i$. 
The goal of sample-pz inference is to infer posterior distributions on parameters $\phi_\text{nz, i}$, utilizing the individual galaxy likelihoods $p(\hat{z}_{\rm p, i} | z_i, \sigma_i, z_{\rm bias, i})$ for all galaxies in the sample.

% This work faciliates a sample selection in latent space. In the context of the presented model this selection would be conducted in terms of intervals (or bins) of true redshift $z$. However, the methodology presented in the following is generic for other quantities of interest. For example we could conduct the same inference with respect to the true colour, magnitude or flux measurement or a combination thereof with true redshift. \lbnotes{this paragraph seems out of place}
\section{Unfolding and Sample Selection}
\label{sec:unfolding_sample_selection}
Many analyses in observational cosmology select galaxies based on apparent photometry or other noisy quantities of interest. In this approach, we select galaxies `with certainty' \citep{2021A&A...647A.124H, 10.1093/mnras/stz2162, 2021MNRAS.505.4249M, 2023MNRAS.524.5109R}, and the uncertainty in the selection propagates to the reconstructed population distribution for said quantity of interest. In the context of weak lensing cluster mass analyses, this means that source redshift distributions can have long tails that can even extend to lower redshifts than the cluster lens. 

The goal of this work is twofold: we want to facilitate a clear selection in the population distribution of quantities of interest, like source sample redshifts, and optimise the selection to reduce biases from \eg\ photometric redshift systematics. This is achieved by unfolding the noisy population distribution onto the latent, noiseless population distribution of quantities of interest. As we will see in the following sections, the selection on an individual galaxy level is then no longer `certain', but each galaxy is attributed a probability, or weight, to be selected. The following sections describe this methodology, starting with a very fast but approximate approach to infer the true, noiseless sample-pz. 
\subsection{Unfolding photometric redshift error}
\label{sub:4_1}
Bayesian analyses in cosmology and astronomy often utilize Markov Chain Monte Carlo (MCMC) techniques for inference. Here the posterior over the parameters of interest is approximated using draws from a Markov chain. MCMC techniques are established as reliable parameter inference techniques. However, they can be limited in scenarios where likelihood evaluations are very costly or where the chains fail to converge to the stationary distribution on acceptable time scales. 

Variational inference is an alternative to MCMC where we make an ansatz for the posterior, the variational distribution $q(\boldsymbol{\phi}_{\rm nz}| \boldsymbol{\alpha})$ in our case, and minimize the Kullback-Leibler divergence $\mathrm{KL}(\dots||\dots)$ \citep{kullback1951information} between this ansatz and the true posterior distribution $p(\boldsymbol{\phi}_{\rm nz}| \mathcal{D})$ of the parameter $\boldsymbol{\phi}$ given the data $\mathcal{D}$: 
\begin{equation}
    q(\boldsymbol{\phi}_{\rm nz}| \hat{\boldsymbol{\alpha}}) = \underset{\boldsymbol{\alpha}}{\argmin} \, \mathrm{KL}\left[q(\boldsymbol{\phi}_{\rm nz}| \boldsymbol{\alpha}) || p(\boldsymbol{\phi}_{\rm nz}| \mathcal{D})\right]
    \label{eq:kl_divergence}
\end{equation}
where $q(\boldsymbol{\phi}_{\rm nz}| \hat{\boldsymbol{\alpha}})$ refers to the variational distribution with optimised parameter $\hat{\boldsymbol{\alpha}}$. 
The result of variational inference is an optimised ansatz that can be used as an approximation to the posterior. 

Minimizing Eq.~\ref{eq:kl_divergence} is equivalent to maximizing the Evidence Lower bound $\mathcal{L}(\boldsymbol{\alpha})$
\begin{equation}
    \mathcal{L}(\boldsymbol{\alpha}) = E_{q(\boldsymbol{\phi_\text{nz}} | \boldsymbol{\alpha})}\left[\log{\left(p(\mathcal{D}, \boldsymbol{\phi_\text{nz}})\right)} - \log{\left(q(\boldsymbol{\phi_\text{nz}} | \boldsymbol{\alpha}) \right)} \right] \, ,
    \label{eq:elbo}
\end{equation}
where we denote the latent variables of the model as $\boldsymbol{\phi_\text{nz}}$ and
$E_{q(\boldsymbol{\phi_\text{nz}} | \boldsymbol{\alpha})}$ denotes the expectation value with respect to samples from the ansatz $q(\boldsymbol{\phi_\text{nz}} | \boldsymbol{\alpha})$.

% , we can write the evidence lower bound\footnote{Also referred to as the negative variational free energy in analogy to the definition of free energy in thermodynamics.} $\mathcal{L}(\boldsymbol{\alpha})$ as:

We see that we have formulated the inference problem in terms of optimising Eq.~\ref{eq:elbo}. In this work, we seek to optimise $\boldsymbol{\alpha}$, such that the variational distribution $q(\boldsymbol{\phi_{\rm nz}} | \boldsymbol{\alpha})$ best describes the true posterior $p(\boldsymbol{\phi_{\rm nz}} | \mathcal{D})$. This approach can be much faster than MCMC, especially in scenarios where we can exploit assumptions on the structure of the variational distribution. In the `mean-field ansatz', the latter is approximated as a product distribution of multiple parameters that are assumed to be independent. This lends itself to a coordinate ascent scheme to optimise the parameters of the variational distribution iteratively. 

In this work we impose a Dirichlet prior on the parameters (\textit{i.e.} the histogram bin heights $\boldsymbol{\phi_\text{nz, i}}$).

The Dirichlet distribution is defined as:

\begin{equation}
    q(\boldsymbol{\phi}_{\rm nz}| \boldsymbol{\alpha}) = \frac{1}{\mathrm{B}(\boldsymbol{\alpha})} \prod_{i = 1}^{N_{\rm bins}} \phi_{\rm i, nz}^{\alpha_i - 1} \, , 
    \label{def:dirichlet}
\end{equation}
where $\sum_{i = 1}^{N_{\rm bins}} \phi_{\rm i, nz} = 1$ and $0 \leq \phi_{\rm i, nz} \leq 1$ and 
\begin{equation}
    \mathrm{B}(\boldsymbol\alpha) = \frac{\prod\limits_{i=1}^{N_{\rm bins}} \Gamma(\alpha_i)}{\Gamma\left(\sum\limits_{i=1}^{N_{\rm bins}}\alpha_i\right)},\qquad\boldsymbol{\alpha}=(\alpha_1,\ldots,\alpha_{N_{\rm bins}}) \, .
\end{equation}
Here $\Gamma(\dots)$ denotes the gamma function.
The Dirichlet distribution is used in many state-of-the-art analysis methodologies \citep{2019arXiv191007127A, 2019MNRAS.483.2801S}, \eg\ in the DES Year 3 analysis \citep{2021MNRAS.505.4249M,  2022MNRAS.510.1223G} because of its simplicity and useful properties for computation. 
The details of the variational inference scheme are detailed in appendix B of \citet{2023MNRAS.524.5109R}. In the following subsection we quote the resulting algorithm. 

\subsubsection{Algorithm: Unfolding of sample-pz}
We obtain the optimised parameter $\hat{\boldsymbol{\alpha}}$ of the variational distribution $q(\boldsymbol{\phi_\text{nz}} | \boldsymbol{\alpha})$ using an iterative scheme. Given a Dirichlet prior with parameter $\boldsymbol{\alpha_0}$, we update the parameter vector of the Dirichlet as: 
\begin{equation}
  \boldsymbol{\alpha}^{t+1} = \boldsymbol{\alpha_0} + \sum_{i = 1}^{N_{\rm gal}} \boldsymbol{\nu_{i}^{t}} \, , 
   \label{eq:Dirichlet_update}
\end{equation}
where $t$ denotes the index of iteration and the sum in Eq.~\ref{eq:Dirichlet_update} runs over the $N_{\rm gal}$-dimension of the $N_{\rm gal}\times N_{\rm bins}$ matrix $\boldsymbol{\nu}$, whose elements are defined as
\begin{equation}
    \nu_{ij}^{t} = \frac{\exp{\left(\psi(\alpha_j^{t}) - \psi(\sum_{a = 1}^{N_{\rm bins}}{\alpha_a^{t}}) + \log{(pz_{\rm ij})}\right)}}{\sum_{k = 1}^{N_{\rm bins}} \exp{\left(\psi(\alpha_k^{t}) - \psi(\sum_{a = 1}^{N_{\rm bins}}{\alpha_a^{t}}) + \log{(pz_{\rm ik})}\right)}} \, .
    \label{eq:var_step2}
\end{equation}
Here $\psi$ denotes the digamma function and the entries $pz_{ij}$ denote the likelihood of the measured photo-z of galaxy $i$ given the redshift integrated over the z-range of the $j$ histogram bin
\begin{equation}
    pz_{ij} = \int  p(\hat{z}_{\rm p, i} | z_i, \sigma_i, z_{\rm bias, i}) \, \mathcal{U}_j(z_i) dz_i \, .
\end{equation}

The elements $\nu_{i j}$ can be interpreted as the `responsibilities' that each histogram bin $j$ has for generating the redshift of galaxy $i$.
Eq.~\ref{eq:Dirichlet_update} and Eq.~\ref{eq:var_step2} are iterated to update the parameter vector of the Dirichlet until convergence. 
\subsection{Sample Selection}
\label{subsec:sample_selection}
We establish a sample selection scheme in latent space by reinterpreting the histogram in Eq.~\ref{eq:sample_pz} as a mixture model consisting of an (included/excluded) sample redshift distribution specified by the parameters $(\boldsymbol{\phi}_{\rm incl}/\boldsymbol{\phi}_{\rm excl})$. Defining the index sets of the (complete/included/excluded) set of bins $\mathcal{B} = \{1, ..., N_{\rm bins}\}$, $\mathcal{B}_{\rm incl} \subseteq \mathcal{B}$, $\mathcal{B}_{\rm excl} = \mathcal{B}\setminus\mathcal{B}_{\rm incl}$ and the weight 
\begin{equation}
    w = 1 - \sum_{i \in \mathcal{B}_{\rm excl}} \phi_{\rm nz, i} \, ,
    \label{eq:w}
\end{equation}
we can introduce the normalized histogram heights of the included
\begin{equation}
    \widetilde{\phi}_{\rm nz, i, incl} = \frac{\phi_{\rm nz, i}}{w} \, ,
    \label{eq:phi_incl}
\end{equation}
and excluded bins
\begin{equation}
      \widetilde{\phi}_{\rm nz, i, excl} = \frac{\phi_{\rm nz, i}}{1-w} \, .
      \label{eq:phi_excl}
\end{equation}
This then allows us to rewrite Eq.~\ref{eq:sample_pz} as: 
\begin{equation}
\begin{split}
    p(z| &\widetilde{\phi}_{\rm nz, i, incl.}, \widetilde{\phi}_{\rm nz, i, excl.}, w ) \\ &= w \left(\sum_{i \in \mathcal{B}_{\rm incl}} \widetilde{\phi}_{\rm nz, i, incl.} \, \mathcal{U}_{i}(z)\right) + (1 - w) \left(\sum_{i \in \mathcal{B}_{\rm excl}} \widetilde{\phi}_{\rm nz, i, excl.} \, \mathcal{U}_{i}(z)\right) \, .
    \end{split}
\end{equation}
We note that the variational distribution over the histogram parameters $\boldsymbol{\phi}_{\rm nz}$ is Dirichlet (see Eq.~\ref{def:dirichlet}). 

By exploiting the properties of the Dirichlet distribution \citep[see][Lemma 1]{Geiger:1970}, we can obtain distributions over the quantities $w$, $ \boldsymbol{\widetilde{\phi}_{\rm nz, incl}}$ and $ \boldsymbol{\widetilde{\phi}_{\rm nz, excl}}$  defined in Eq.~\ref{eq:w}, Eq.~\ref{eq:phi_incl} and Eq.~\ref{eq:phi_excl} respectively. Using the aggregation property of the Dirichlet we obtain
\begin{equation}
    w \sim \mathrm{Beta}\left(\overline{\alpha}_{\rm incl}, \overline{\alpha}_{\rm excl} \right) \, , 
\end{equation}
where we defined $\overline{\alpha}_{\rm incl} = \sum_{i \in \mathcal{B}_{\rm incl}} \alpha_i$ and $\overline{\alpha}_{\rm excl} = \sum_{i \in \mathcal{B}_{\rm excl}} \alpha_i$.

The probability density function of the beta distribution is defined as: 
\begin{equation}
    p(w | \overline{\alpha}_{\rm incl}, \overline{\alpha}_{\rm excl}) = \frac{\Gamma(\sum_{i = 1}^{N_{\rm bins}} \alpha_i)}{\Gamma(\overline{\alpha}_{\rm incl})\Gamma(\overline{\alpha}_{\rm excl})}\, w^{\overline{\alpha}_{\rm incl}-1}\left(1-w\right)^{\overline{\alpha}_{\rm excl}-1} \, .
\end{equation}
We further note that for $(m/n) \in (\mathcal{B}_{\rm incl}/\mathcal{B}_{\rm excl})$ we have: 
\begin{equation}
\begin{split}
    \boldsymbol{\widetilde{\phi}_{\rm nz, incl}} &\sim \mathrm{Dir}(\{\alpha_m\}) \\
    \boldsymbol{\widetilde{\phi}_{\rm nz, excl}} &\sim \mathrm{Dir}(\{\alpha_n\})
    \end{split}
    \label{eq:included_excluded_phi}
\end{equation}

Since both $\boldsymbol{\widetilde{\phi}_{\rm nz, excl}}$ and $\boldsymbol{\widetilde{\phi}_{\rm nz, incl}}$ are independent parameter sets that each lie on their corresponding probability simplex and the construction of the aforementioned mixture is straightforward, we can construct this mixture model for all possible combinations of (included/excluded) bins. 

This allows us to efficiently optimise the partition of bins into included/excluded groups using an optimisation scheme that takes into account the expected systematic bias incurred by a potential model misspecification error from ill-calibrated galaxy redshifts and the resulting loss in constraining power. After optimisation, we will only use the included histogram bins in a subsequent weak lensing cluster mass measurement analysis.
\subsection{A physical interpretation of the Dirichlet distribution}
In the following we provide a physical interpretation of the Dirichlet in the context of this work. Let $N_i$ denote the number of galaxies in a redshift bin $i$. We further assume that $N_i$ is Poisson distributed with intensity parameter $\tau_i$ that itself is distributed according to a restricted gamma distribution:
\begin{align}
    \tau_i &\sim \mathrm{Gamma}(\alpha_i, \beta=1)\\
    N_i &\sim \mathrm{Pois}(\tau_i)\, .
\end{align}
The Poisson distribution can be further decomposed into the product of a multinomial of the relative counts $\phi_i$ and the Poisson distribution of the total counts $N_{\rm gal}$, so we can equivalently write:
\begin{align}
    \tau_i &\sim \mathrm{Gamma}(\alpha_i, \beta=1)\\
    N_{\rm gal} &\sim \mathrm{Pois}(\tau_{\rm tot}) \\
    N_i &\sim \mathrm{Mult}(\boldsymbol{\phi}, N_{\rm gal}) \, , 
\end{align}
where $\tau_{\rm tot}$ and $\boldsymbol{\phi}$ are given as:
\begin{equation}
    \begin{split}
        \tau_{\rm tot} &= \sum_{i=1}^{N_{\rm bins}} \tau_i \\
        \phi_i &= \frac{\tau_i}{\sum_j^{N_{\rm bins}} \tau_j} \qquad 1 \leq i \leq N_{\rm bins}
    \end{split}
\end{equation}
The total number of galaxies is fixed in a typical cosmological analysis, so the only parameter vector that's relevant here is $\boldsymbol{\phi}$. It can be shown that if $\boldsymbol{\phi}$ is sampled according to
\begin{align}
        \tau_i &\sim \mathrm{Gamma}(\alpha_i, \beta=1)\\
        \phi_i &= \frac{\tau_i}{\sum_j^{N_{\rm bins}} \tau_j} \qquad 1 \leq i \leq N_{\rm bins} 
\end{align}
it is distributed according to a Dirichlet with parameter vector $\boldsymbol{\alpha}$. Thus a Dirichlet  prior on $\boldsymbol{\phi}$ can be interpreted as a gamma prior on the galaxy count rates along the line-of-sight. This can lead to a restricted treatment of between-bin correlations as discussed in \citet{2023MNRAS.524.5109R}. Nonetheless, using the Dirichlet can be very practical in cases where the likelihood dominates the inference, and computational speed is paramount. These advantages motivate the Dirichlet in this work. 

\subsection{Data}
\begin{figure}
    \centering
    \includegraphics[scale=0.55]{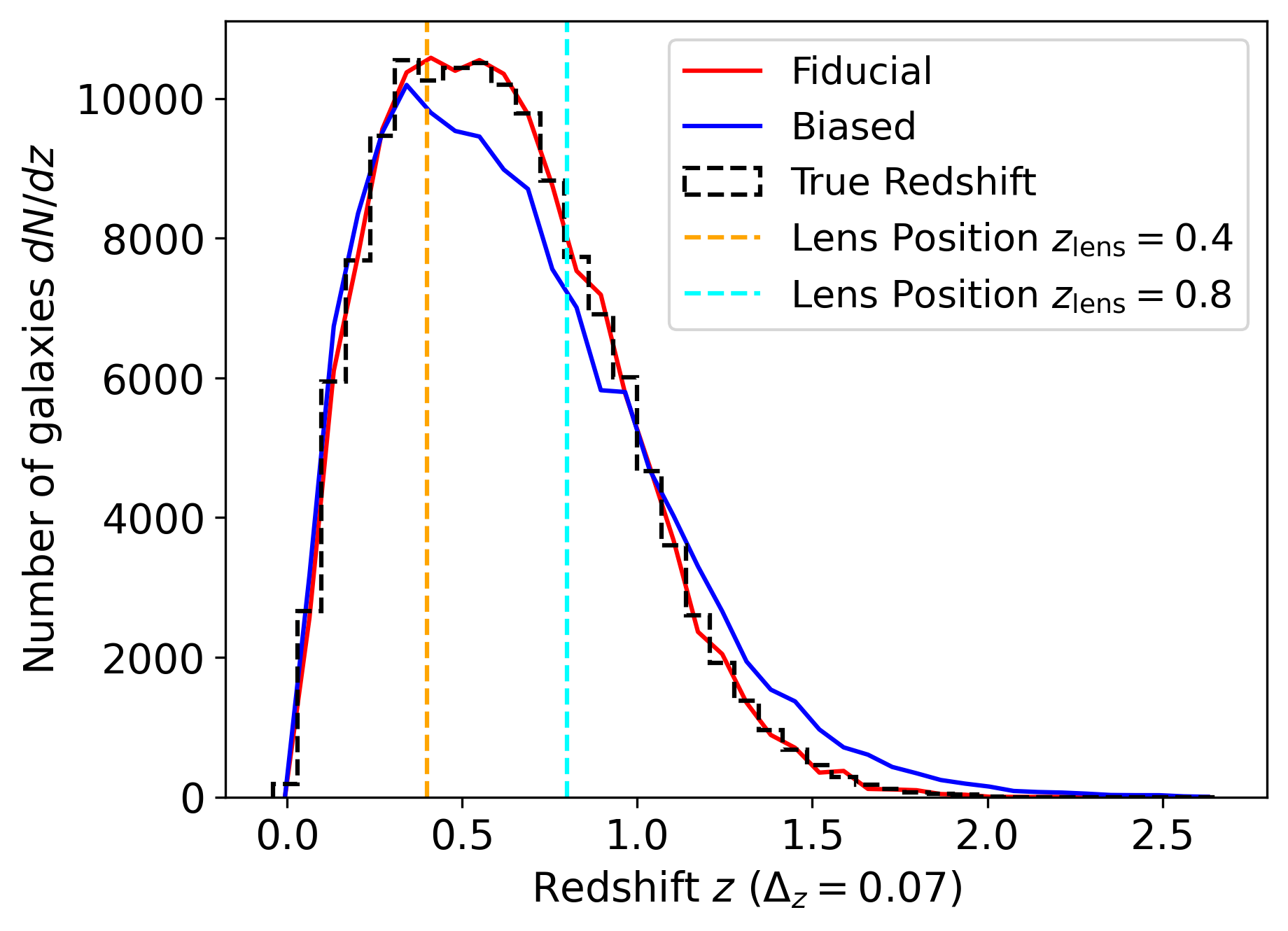}
    \caption{ Histogram of the number of galaxies ($dN/dz$) as a function of redshift $z$ in bins of size $\Delta_z = 0.07$. The black histogram shows the distribution of true redshifts, the vertical (orange/cyan) dashed line the redshift of the cluster lens at $z_{\rm lens}=(0.4/0.8)$. The (red/blue) lines show the sample redshift distribution estimates obtained for the (Fiducial/Biased) scenarios as described in the text.  }
    \label{fig:1}
\end{figure}

As data for our mock study, we use the full catalog of ``target'' data generated by \citet{2020MNRAS.496.4769H} that mimics a Dark Energy Survey Year 1 sample redshift distribution. The sample contains 134,155 galaxies with a median i-band magnitude of 23. We show the sample redshift distribution in Fig.~\ref{fig:1} here in terms of the number counts ($dN/dz$) as a function of redshift $z$ in histogram bins of size $\Delta_z = 0.07$. The black histogram shows the distribution of true redshifts and the vertical (orange/cyan) dashed line denotes the redshift of our example galaxy cluster lens at ($z_{\rm lens} = 0.4$/$z_{\rm lens} = 0.8$). The (red/blue) lines show the point estimates obtained for the (Fiducial/Biased) scenarios, derived by running the aforementioned variational inference scheme to obtain the posterior mean estimate of the histogram heights $\boldsymbol{\hat{\phi}_\text{\rm nz}}$ multiplied by $N_{\rm samp}$. 
% \textbf{todo: Remember to redo with the +1 added, makes no visible difference but check, change plot label of fig1 to Fiducial}

\section{Sample Selection optimisation}
\label{sec:4}
\S~\ref{sec:unfolding_sample_selection} introduced our fast methodology to unfold source sample redshift distributions (sample-pz) and our sample selection methodology that partitions the source sample into included and excluded populations using a mixture model. In the following section we will discuss our methodology to optimise this selection, balancing the tradeoff between removing bins with a high probability of containing galaxies with biased redshifts and keeping a larger source sample for increased constraining power on the inferred cluster mass. We will start the discussion by taking a look at suitable metrics to measure the aforementioned tradeoff.

%\lbnotes{this paragraph seems out of place}
\subsection{Metrics}
\label{subsec:metrics}
As discussed in \S~\ref{sec:3} it is appropriate to quantify photometric redshift error induced systematics in weak lensing cluster mass estimates using the difference between the true $\Sigma_{\rm \crit, true}$ and a critical surface mass obtained using an incorrect, i.e. misspecified, model for the sample redshift distribution $\Sigma_{\rm \crit, estim}$. In the following, we make the simplifying assumption that we can estimate this systematic using the photometric redshift calibration techniques described in \S~\ref{sec:1}. This means that we are assuming that we are able to use \eg\ spatial cross-correlation redshift estimation techniques to calibrate the sample-pz and we neglect the intrinsic error or the residual systematics in this calibration technique. This will not be perfectly possible in practice, and we will discuss the effect of calibration error on the presented optimisation scheme in \S~\ref{sec:discussion}.   

Since $\Sigma_\crit$ is an integrated quantity we note that an underestimation/overestimation of the integrand as a function of redshift can be compensated by an overestimation/underestimation at a different redshift. A comparable example of this is illustrated in Fig.~19 of \citet{2015MNRAS.452.3710R} in the context of the lensing efficiency term in cosmic shear power spectra that can ``hide'' model error due to photometric redshift systematics. Thus, the value of $\Sigma_{\rm crit}$, being a quantity integrated over the sample-pz, might not be globally biased, because of compensating model misspecifications in the integrand. This is still undesirable, because the resulting model misspecification will depend on the lensing geometry, impact inferences of cosmological parameters that enter the integrand and mis-calibrate uncertainty quantification. We, therefore, only consider metrics that are defined in redshift intervals of the source sample and provide summary statistics that are robust against this effect. In the following we denote the corresponding quantities of interest with the subscript (fid/bias) for (fiducial/biased).
%\lbnotes{Point out explictly this is unstable/dangerous?}
% The relative bias cost 
% \begin{equation}
%     \mathcal{C}_{\rm rel} = \sum_{i = 1}^{N_{\rm bins}} \left|\frac{\Sigma_{\rm crit, i, fid} - \Sigma_{\rm crit, i, bias}}{\Sigma_{\rm crit, i, fid}}\right| = \sum_{i = 1}^{N_{\rm bins}} \left|\frac{\phi_{\rm nz, i, fid} - \phi_{\rm nz, i, bias}}{\phi_{\rm nz, i, fid}} \right|  \, ,
%     \label{eq:crel}
% \end{equation}
% where have used Eq.~\ref{eq:sig_crit} and Eq.~\ref{eq:sample_pz}. 

The absolute bin-wise distance between biased and fiducial $\Sigma_\crit$ estimates in bin $i$ is defined as
\begin{equation}
            \mathcal{C}_{2}(\mathcal{B}) = \ \sum_{i \in \mathcal{B}} \left| \Sigma_{\rm \crit, i, fid} - \Sigma_{\rm \crit, i, bias}\right| \, , 
    \label{eq:c2}
\end{equation}
where $\mathcal{B}$ denotes the index set of the (fiducial/biased) histogram bins and 
\begin{equation}
    \Sigma_{\rm \crit, i} =  \phi_i \, \Sigma_{\rm \crit, i, mid} \, .
\end{equation}
Here, $\Sigma_{\rm \crit, i, mid}$ denotes the integrand of Eq.~\ref{eq:sig_crit} evaluated at the midpoint of the $i$ histogram bin and $\phi_i$ the normalized `height' of sample-pz bin $i$ (see Eq.~\ref{eq:sample_pz}). We note that we will only consider sample redshift distribution point estimates, which we define using the mean of the Dirichlet posterior $p(\boldsymbol{\phi}_{\rm incl} | \boldsymbol{\alpha}_{\rm incl}, \mathcal{D})$ and quantities derived thereof in the following until otherwise stated, \ie\ we are not considering posterior uncertainty in the sample redshift inference. 

% These cost functions will be used in the following sections to optimise the latent space to minimize the impact of photometric redshift systematics on weak lensing cluster mass measurements. 

\subsection{Combinatorial optimisation}
We define the ``total gain'' $\mathcal{T}$ as the tradeoff between the ``value'' of the selection $\mathcal{V}$ and the summed, or ``total cost'' $\mathcal{C}$ of all included redshift bins in terms of their impact on systematic biases in $\Sigma_\crit$
\begin{equation}
    \mathcal{T} = \lambda \mathcal{V} - (1 - \lambda) \, \mathcal{C} \, ,
    \label{eq:total_loss}
\end{equation}
where $\lambda$ is the tradeoff parameter (see below). 
Using Eq.~\ref{eq:c2} we define the cost function $\mathcal{C}$ as:
\begin{equation}
            \mathcal{C} = \frac{\mathcal{C}_{2}(\mathcal{B}_{\rm incl})}{\mathcal{C}_{2}(\mathcal{B}_{>z_{\rm lens}})}
    \label{eq:def_cost}
\end{equation}
and the value function $\mathcal{V}$ as
\begin{equation}
    \mathcal{V}(\mathcal{B}_{\rm incl}) = \frac{\sum_{i \in \mathcal{B}_{\rm incl}} N_{i}}{ \sum_{j \in \mathcal{B}_{>z_{\rm lens}}} N_{j}} \, .
\end{equation}
Here, $N_{i}$ denotes the number of galaxies in redshift histogram bin $i$ and the index set $ \mathcal{B}_{>z_{\rm lens}}$ denotes all sample-pz histogram bins with midpoint redshift larger than $z_{\rm lens}$. 
Our optimisation objective is to find $\mathcal{B}_{\rm incl} \in \mathcal{P} (\mathcal{B})$\footnote{Here, $\mathcal{P}(\mathcal{B})$ denotes the powerset of $\mathcal{B}$.} such that $\mathcal{T}(\mathcal{B}_{\rm incl})$ is maximized for a selected value of $\lambda$. Maximizing $\mathcal{T}$ implies finding the maximum sample size that we retain while minimizing the bias in the modelling of $\Sigma_{\rm crit}$. The tradeoff between these terms is governed by the $\lambda$ variable.
% \begin{gather}
% \begin{split}
%     \underset{\mathcal{B}_{\rm incl}}{\max}\,  &V(\mathcal{B}_{\rm incl}) \\
%     {\rm while} \ &\mathcal{C}(\mathcal{B}_{\rm incl}) \leq T \, .
%     \end{split}
%     \label{eq:optim_obj}
% \end{gather}
% Here $\mathcal{C}$ can denote either $\mathcal{C}_{\rm rel}$ or $\mathcal{C}_{2}$. The value function chosen in this work is the total number of galaxies in the included redshift bins:
% \begin{equation}
%     V(\mathcal{B}_{\rm incl}) = \sum_{i \in \mathcal{B}_{\rm incl}} \phi_{\rm fid, i} \cdot N_{\rm samp} \, .
%     \label{eq:v}
% \end{equation}
% Eq.~\ref{eq:v} is clearly a very simplistic value function and wee will discuss in \S~\ref{sec:5} various extensions. 

The problem described in this section resembles a knapsack problem, where we would like to assemble bins such that the residual systematic is below a given threshold while maximizing \eg\ the amount of background galaxies kept in the sample, thus increasing the signal-to-noise. However, a naive application of the knapsack optimisation using dynamic programming is not valid because we optimise the construction of a mixture model. In this case, it can be shown by counterexample that dynamic programming, in general, will not find the optimal solution. 

% Concretely, after each application of the \textit{do(x)} operator we change the model \markus{by fixing a given redshift bin to zero and thus} the restriction $\sum_{i \in \mathcal{B}_{\rm incl}} \boldsymbol{\phi}_{i} = 0$ implies that the selected bins are elements of a ``new'' simplex. This is also true for other loss functions, \eg\\ the relative difference between biased and fiducial redshift distribution bin heights. 
%\lbnotes{because we are summing over different bins? Can you be more clear?} 
%, which can be easily proven by counterexample\footnote{Consider for example a 3-bin case with: $\boldsymbol{\phi}_{\rm fid} = (0.01, 0.09, 0.9)$, $\boldsymbol{\phi}_{\rm bias} = (0.1, 0.9, 0.8)$, with $N_{\rm samp} = 6$.   }.
\subsubsection{Algorithm: Cross-Entropy optimisation}
\label{sec:4_3_1}
A flexible and easy method to perform combinatorial optimisation that we found to work well in this context is cross entropy optimisation \citep{rubinstein1999, BOTEV201335}.  This probabilistic optimisation scheme starts by defining an initial probability vector of Bernoulli variables with $p_i = 0.5$ for all $i \in \mathcal{B}$. 

We then repeat the following steps until convergence:  
\begin{enumerate}
	\item Generate 1000 realisations of the multivariate Bernoulli distribution
	\item Evaluate $\mathcal{T}$ on each generation
	\item Select the best 50\% realisations with respect to $\mathcal{T}$
 \item Update $p_i = \sum_{j=1}^{N_{\rm selected}} \mathbf{1}_{i j}$, where the sum goes over the number of selected realisations $N_{\rm selected}$ and $\mathbf{1}_{i j}$ denotes the indicator function that is unity for each realisation $j$, if $i \in \mathcal{B}_{\rm incl}$ and zero otherwise
 \item We continue with (i) using the updated $\boldsymbol{p}$.
\end{enumerate}
 \begin{figure*}
    \centering
    \includegraphics[scale=0.3]{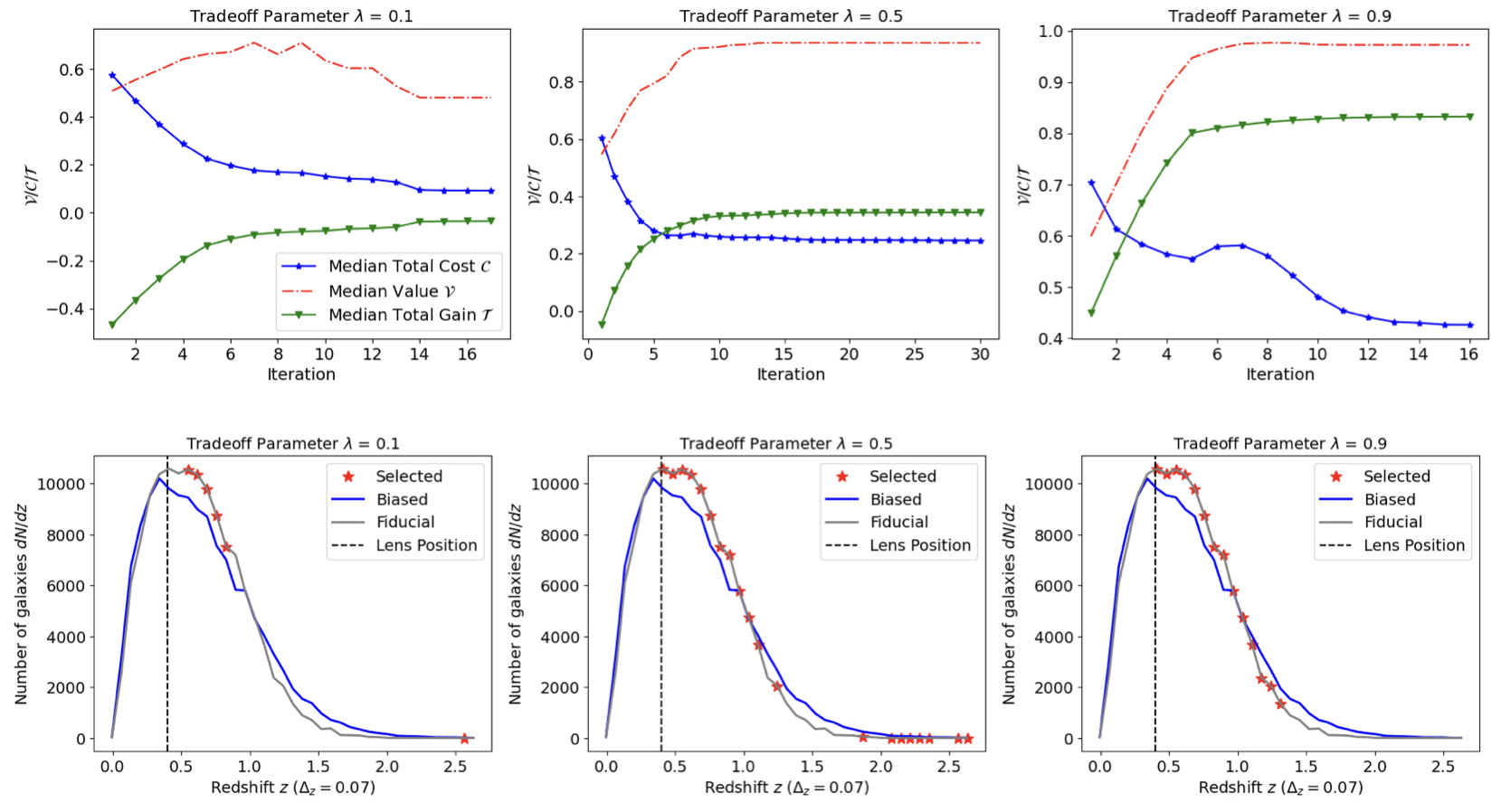}
    \caption{ Combinatorial optimisation results for cluster lens redshift of $z_{\rm lens} = 0.4$. \textit{Upper panels:} The median Cost $\mathcal{C}$ (blue, stars), Value $\mathcal{V}$ (red, doted-dashed) and Total gain $\mathcal{T}$ (green, triangle) as a function of the Cross Entropy optimisation iteration for different values of the tradeoff parameter $\lambda$. \textit{Lower panels:} Redshift bins selected by combinatorial optimisation for different tradeoff parameters $\lambda \in [0, 1]$. The algorithm discussed in \S~\ref{sec:4} (removes/retains) more bins for (smaller/larger) values of $\lambda$. The (grey/blue) distribution shows the (Fiducial/Biased) redshift distributions for a redshift binning of $\Delta z = 0.07$. The vertical black dashed line represents the redshift of the cluster lens and the red stars indicate the selected bins after optimisation. The optimisation is performed using the Cross Entropy method using the $\mathcal{C}_2$ loss function (Eq.~\ref{eq:c2}) as described in \S~\ref{sec:5}.  }
    \label{fig:2}
\end{figure*}
 \begin{figure*}
    \centering
    \includegraphics[scale=0.28]{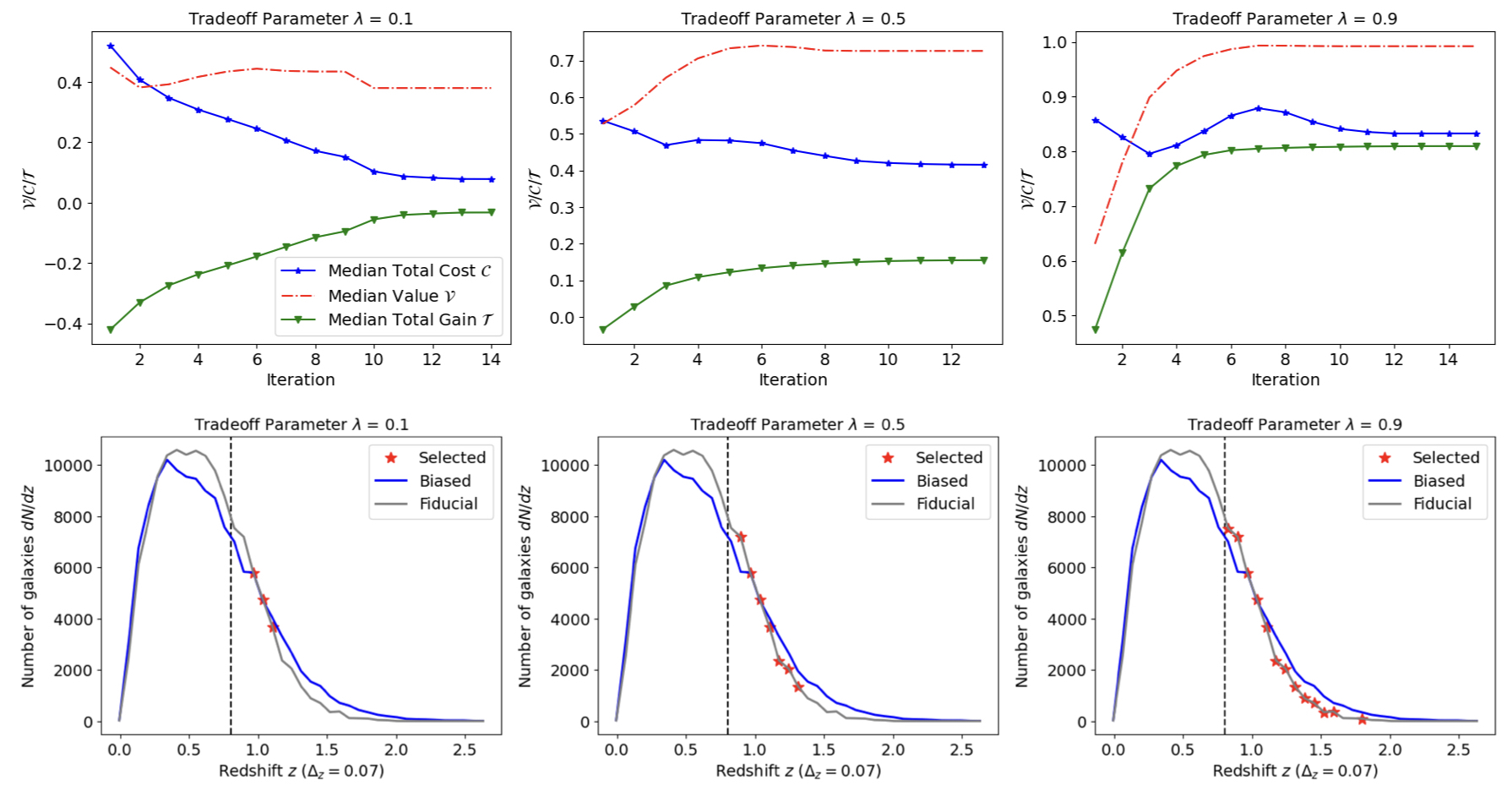}
    \caption{ Combinatorial optimisation results for cluster lens redshift of $z_{\rm lens} = 0.8$. Panel description is the same as in Fig.~\ref{fig:2}. }
    \label{fig:3}
\end{figure*}

% As a result a naive application of a knapsack optimisation that treats the pointwise terms in $\mathcal{C}_{2}$ as separable will lead to incorrect solutions, as can be easily proven by counterexample

\section{Results}
\label{sec:5}
We demonstrate the optimisation (see \S~\ref{sec:4}) using the photometric redshift error model (see \S~\ref{sec:2}) for parameters listed in Tab.~\ref{tab:2} and a cluster lens at redshift $z_{\rm lens} = 0.4$ (see Fig.~\ref{fig:2}) and $z_{\rm lens} = 0.8$ (see Fig.~\ref{fig:3}). 

We run the algorithm for 10 iterations of steps (i) - (v) with 1000 realisations of the Bernoulli, cutting at the 0.5 quantile to select the best samples from each set of realisations using the total gain $\mathcal{T}$ as the optimisation objective. 

The upper panels of Fig.~\ref{fig:2} and Fig.~\ref{fig:3} show the tradeoff between the value and cost, where we show the median cost $\mathcal{C}$ (blue, stars), value $\mathcal{V}$ (red, doted-dashed) and total gain $\mathcal{T}$ (green, triangle) as a function of the cross entropy optimisation iteration. The subpanels correspond to different values of the tradeoff parameter $\lambda$. As expected we see that the total gain is dominated by the cost function for low $\lambda$, converging within 10 iterations. In this case fewer galaxies are retained in the sample. Similarly, higher values of $\lambda$ put less weight on the reduction of systematic bias due to photometric redshift error and, therefore, the final sample retains more galaxies. We see in all panels that the scheme converges within less than 30 iterations, often after 10. Since we are using a heuristic combinatorial optimisation scheme, the $\mathcal{C}$, $\mathcal{V}$ and $\mathcal{T}$ don't have to monotonially decrease, even though this is true in the vast majority of cases. For $\lambda = 0.5$, we reduce the critical surface mass density $\Sigma_\crit$ modelling bias, as measured using the $\mathcal{C}$ metric, on the 60-70\% level, while maintaining up to 90\% of galaxies. More extreme reductions in bias can be achieved using \eg\ $\lambda = 0.1$, where a reduction of over 90\% in the model misspecification $\mathcal{C}$ can be achieved while still maintaining around $50\%$ of galaxies.

The lower panels of Fig.~\ref{fig:2} and Fig.~\ref{fig:3} show the (Fiducial/Biased) redshift distributions in (grey/blue). We select a redshift bin size of $\Delta z = 0.07$ and indicate the respective cluster lens redshift using the vertical black dashed line. The red points indicate the selected bins after optimisation. The different subpanels show the results for different tradeoff parameters $\lambda \in [0, 1]$. As discussed in \S~\ref{sec:4} the algorithm (removes/retains) more bins for (smaller/larger) values of $\lambda$. As a result the selected redshift bins (red stars) are more numerous and extend to higher redshift for larger values of $\lambda$. Note that $\lambda$ should not be interpreted as a fraction because it represents a relative weight between a value and total cost function. While the value function in this case is the total number of galaxies in the sample, the relative fraction of galaxies retained after optimisation will not equal $\lambda$.  

We identify a bias between fiducial and biased sample redshift distributions at $z \in [0.5, 1.0]$, which is close to the cluster lens for both lens redshifts. Despite this apparent bias, the combinatorial optimisation scheme consistently selects redshift bins in this redshift region. This is because Eq.~\ref{eq:sig_crit} is an increasing function of redshift, which means that lower redshift bins will contribute less to the overall cost. At the same time, we note that the redshift bins closer to the lens also contain the majority of galaxies in the source sample, which incentivises retaining them. In contrast, we note that the tail distribution is consistently not selected. Here it is worth noting that the tail bins in the $(\lambda = 0.1;\;z_{\rm lens} = 0.4)$ and $(\lambda = 0.5;\;z_{\rm lens} = 0.4)$ cases are selected by the algorithm. These bins have very little contribution to the total gain and would not be relevant in a real analysis. Given that we use a stochastic algorithm for the combinatorial optimization it is expected that these bins can be included due to the intrinsic stochasticity in the algorithm. These local minima will be more pronounced in scenarios where the impact of the cost function is high. This nominally favors the inclusion of these tail bins due to almost identical histogram heights compared with the fiducial case.

We note however, that a more complex modelling, that includes the redshift-dependent signal-to-noise in the tangential shear measurement into the analysis will likely favour a different optimal bin configuration. We leave this investigation for future work. We finally note that computationally the cross-entropy optimisation is quite efficient, where convergence is obtained within 30 seconds on a Macbook Air M2 2022. This low computational cost, coupled with the modelling flexibility supported by this very generic optimisation routine, makes this simple method an ideal choice for the more complex modelling approaches that we want to tackle in the future.  

% \lbnotes{need to add the quantitative numbers \eg\, Cost here, and reduction in uncertainty in sigma crit for different scenarios}
\section{Practical Application to Cluster Mass Measurements}
\label{sec:discussion}
%Explain how the algorithm integrates into a real WL galaxy cluster analysis. 
In the following, we detail how the described approach to sample selection can be applied to a practical cluster weak-lensing analysis. We discuss the prospective application in 5 steps: defining the optimisation objective and hierarchical population model, fitting the hierarchical population model using the full data, selecting a suitable decomposition in included and excluded populations, running the optimisation, extracting the results and using the optimised galaxy sample in the subsequent analysis. 
\subsection{Optimisation objective and hierarchical population model}
Starting with the analysis, we define a suitable cost and value function. The cross-entropy optimisation algorithm is used here to optimise a total gain function that is constructed from the tradeoff between expected cost and value. Alternatively, if we want the cost of the optimised sample to lie below a certain limit, the cross-entropy optimisation algorithm can also optimise the value function while maintaining a cost below a set threshold. This flexibility allows us to tune the sample optimisation towards specific science goals, such as reducing the systematic error budget below the statistical.

We then define and fit a statistical model to describe the population of galaxies in latent space. In this work, we use variational inference for good performance during combinatorial optimisation. However, other techniques like MCMC methods are also possible. In fact, while variational inference is a useful framework to conduct inference during optimization, we would still recommend running an MCMC chain on the final, optimized, selection, as we expect techniques like Hamiltonian MCMC to have better error quantification. We reiterate that this is of less concern while optimizing the tradeoff between systematic uncertainty and sample size where the maximum a-posteriori prediction is adequate to quantify systematic biases. For further details on the inference methodology we refer the reader to \citet{2022MNRAS.509.4886R, 2023MNRAS.524.5109R}.

\subsection{Sample Selection}
After defining a suitable optimisation objective and fitting a hierarchical population model that describes the distribution of galaxy quantities of interest in latent space, we can define the bins that are included and excluded in the population model. Since we only use the included latent parameters of the population model in the subsequent inference, it effectively removes the influence of galaxies with a high probability of lying within excluded bins from the fit. Thus, if their photometric redshift is incorrect, we can significantly remove biases in the modelling of the critical surface density in this way. 
\subsection{Running the optimisation}
The cross-entropy optimisation scheme discussed in this work provides a flexible and fast approach to optimise the selection of included and excluded bins to optimise the total gain. The cross-entropy algorithm was described in \S~\ref{sec:4_3_1}, where the tuning parameters are the total number of realisations drawn from the multivariate Bernoulli distribution in step (i) and the fraction of realisations removed at each iteration step in (iii). Increasing the number of realisations generated by the multivariate Bernoulli distribution will lead to a smoother convergence and overall better optima at the expense of a higher computational cost. For this work we chose 1000 realisations as a good tradeoff between these considerations. The fraction of removed realisations is related to the total number of realisations. Removing a higher fraction of generated realisations leads to faster convergence at an increased risk of getting stuck in a local minimum. In our case, we chose 50\% in step (iii) as a good tradeoff. We note that since we are assuming the usage of spatial cross-correlations to calibrate the sample redshift distribution and will obtain error estimates that enter our cost functions, we need to ensure that the cross-correlation data vector takes these selections into account. Since our selection is in redshift and all common cross-correlation techniques constrain the sample redshift distribution in redshift bins, there are no further steps required for our current setup. However, if the selection would be in other quantities of interest like colour, or parameters like stellar mass or star formation rate, we would need to construct additional cross-correlation measurements for each bin in this latent space. We, therefore, recommend using cross-correlation redshift inference codes similar to \textit{The-wizz} \citep{Morrison2016} that support the inclusion of these selection functions without re-evaluating the pair counts.
% \subsection{Extracting the results}
% The results of the optimisation include a set of bins of the population distribution that are (included/excluded) from the final sample and the corresponding posterior weights that describe the probability that a galaxy is (included/excluded). We use posterior mean point estimates for these quantities derived from the full posterior, which are a direct byproduct of fitting and optimizing our hierarchical population model.

\subsection{Cluster mass inference}
\label{sec:illustrative_mock}
This section will illustrate how an optimized galaxy sample can be used in a subsequent analysis and how minimizing the cost function in Eq.~(\ref{eq:def_cost}) enables a better control of the systematic error in the resulting cluster mass measurement. We want to illustrate our methodology using a mock catalog of redshifts and tangential shear measurements for a cluster lens at redshift $z_{\rm Lens} = 0.5$. We consider a population of photometric outliers at $z=2.5$ sampled with a standard deviation of $\sigma = 0.3$ constituting 10\% of the sample motivated by Rubin forecasts \citep{2018AJ....155....1G}. This localization of the redshift failures in the sample redshift distribution is common and allows for easy visual interpretation in the context of this illustrative example. We provide more details on the creation of this mock catalog in Appendix~\ref{sec:mock_sim}.

\begin{figure}
    \centering
    \includegraphics[scale=0.5]{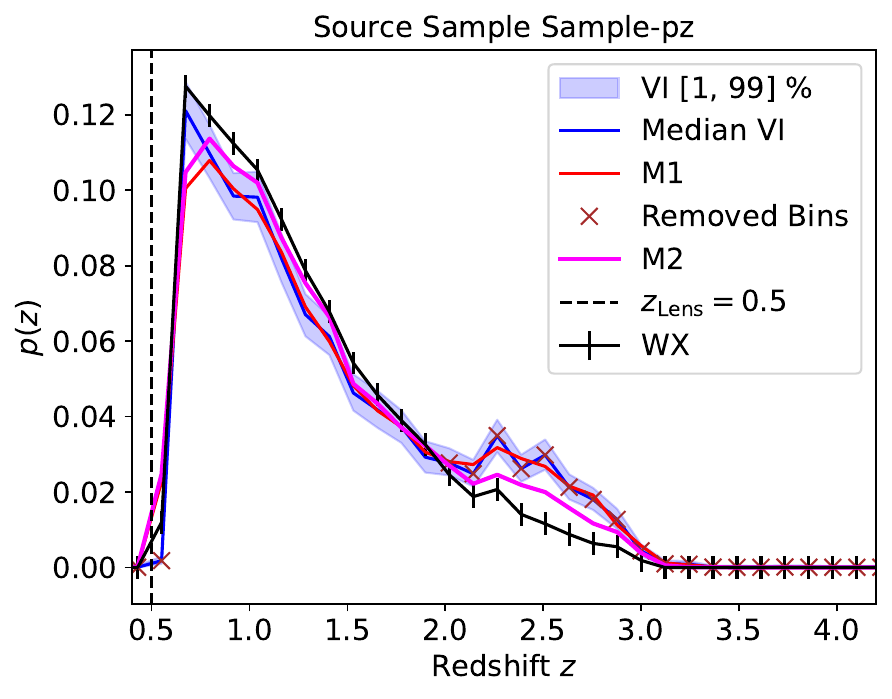}
    \caption{ True source sample redshift distribution with errorbars illustrating a cross-correlation redshift calibration measurement (black solid, `WX'), the cluster redshift ($z_{\rm Lens}$), the results from two photometric redshift methods M1 and M2 and the posterior (i.e. unfolded) source sample redshift distribution ($[1, 99]\%$ percentiles 'VI [1, 99]\%', median 'Median VI').  The histogram is normalized such that the histogram bins sum to unity. }
    \label{fig:mock_study_illustration}
\end{figure}

Our simplified cluster mass measurement uses the following data products which are standard in many modern cluster mass analyses:
\begin{enumerate}
    \item photometric galaxy source catalog
    \item results from two photometric redshift algorithms (`M1' and `M2') that produce (finite accuracy) estimates of the photometric redshift of the source galaxies. 
    \item measurements of the spatial cross-correlation between a reference spectroscopic sample like DESI and the photometric catalog (`WX') 
    \item a shape catalog with tangential shear measurements.
\end{enumerate}
We show these different data products in Fig.~\ref{fig:mock_study_illustration} and refer to Appendix \ref{sec:mock_sim} for further details. The figure shows the true source sample redshift distribution with errorbars resembling a cross-correlation measurement (black solid, `WX'), the cluster redshift ($z_{\rm cluster}$), the results from M1 and M2, both of which are noisy estimates of redshift as well as the posterior of the unfolded source sample redshift distribution based on M1 ($[1, 99]\%$ percentiles `VI [1, 99]\%', median `Median VI').

When applying the combinatorial optimization we can now select a fiducial method to compare M1 against. In our setup this could be either M2 or WX, which would then be used to calculate $\Sigma_{\rm fid}$ in Eq.~\ref{eq:def_cost}. In practise we want to select the method that we deem to be most independent from M1. This would be WX as it is based on spatial clustering rather than photometry. We can then commence the optimization wrt. Eq.~\ref{eq:total_loss}. This is described throughout this paper; we focus here on how to use the results of a sample optimization. We reiterate that the goal of our methodology is to localize potential model misspecification by treating the difference between M1 and WX as a conservative lower bound of its impact on cluster mass estimation, which we quantify by Eq.~\ref{eq:def_cost}.

Consider that the outcome of an optimization is indicated as the brown crosses (`Removed Bins'). For this selection we yield a relative reduction in $\mathcal{C}$ of
\begin{equation}
    \frac{\mathcal{C}_{\rm orig} - \mathcal{C}_{\rm optim}}{\mathcal{C}_{\rm orig}} = 0.40 \, ,
\end{equation}
where $\mathcal{C}_{\rm orig}$ refers to the result we obtain when we keep all the bins and $\mathcal{C}_{\rm optim}$ the result we obtain when we remove the marked bins. The reduction in sample size is 20\% where we run the optimization with $\lambda = 0.5$. We also removed two bins close to the cluster lens position because the size of the bins might extend to redshifts smaller than the cluster lens. This however has negligible impact on the total gain. We note that especially for well localized outliers like the scenario shown in Fig.~\ref{fig:mock_study_illustration} visual inspection and physical intuition about the cause of outliers will be valuable to interpret and tune the combinatorial optimization. This is a vital step since combinatorial optimization schemes are in general not guaranteed to find global optima and should not be used as a `black-box'.  Irrespective of the optimization scheme we stress that running the deconvolution and selection methodology given a choice of bins is always necessary. We further note that in order to obtain high-quality sample redshift inference with well calibrated uncertainties, it is vital to include sample redshift information from the galaxies photometry, and it will not be possible to rely on a single element of the model set, for example WX. 

We next check that the optimization result resembles the model misspecification error we see across all elements of our models set. Our model set includes in this example M1, M2 and WX. Comparing with M2 we clearly see that both WX and M2 deviate consistently within the region where we would expect outlier populations, confirming the selection.

During optimization we needed to balance the reduction in the loss Eq.~\ref{eq:def_cost}  with the reduction in sample size. The choice of this tradeoff is a tuning parameter that can be adjusted in practise by forecasting the impact on the subsequent analysis in a sensitivity study. In the presented case of localized outliers the choice of $\lambda$ is not critical for the results, however in the more complex scenario presented in the previous sections it will be a vital tuning parameter. 

\begin{figure}
    \centering
    \includegraphics[scale=0.5]{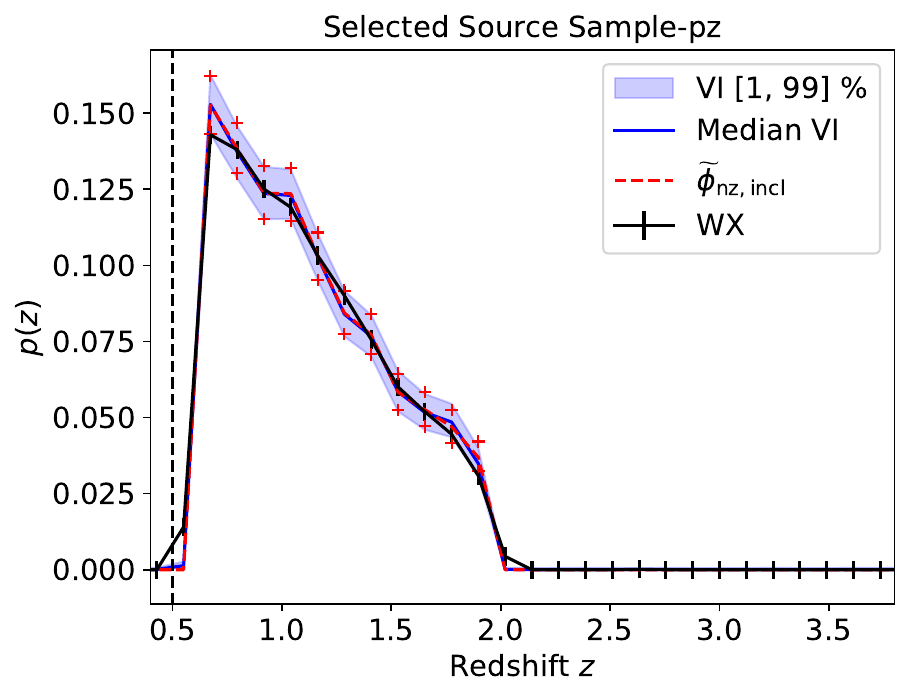}
    \caption{ Source Sample redshift distribution after selection:  (`VI [1, 99]\%', `Median VI') denote the corresponding percentile intervals and the median obtained by rerunning the deconvolution algorithm directly on the sample. We remove galaxies that fall within the `Removed Bins' shown in Fig.~\ref{fig:mock_study_illustration}. The dashed red (median) and red points ([1, 99]\% percentile interval) denote the corresponding result from Eq.~\ref{eq:included_excluded_phi}, `WX' denotes the true sample redshift distribution with cross-correlation measurements shown as errorbars. The histogram is normalized such that the histogram bins sum to unity. }
    \label{fig:redshift_selection_removal}
\end{figure}

We can now derive weights using the probability that the redshift of galaxy $i$ is within the included bins, denoted as $\omega_i$, 
\begin{equation}
    \omega_i = \frac{\sum_{a \in \mathcal{B}_{\rm incl}}\hat{\alpha}_a  pz_{i a} }{\sum_{a \in \mathcal{B}} \hat{\alpha}_{a}  pz_{i a} } \, ,
    \label{eq:weights}
\end{equation}
where $\boldsymbol{\hat{\alpha}}$ is the parameter vector of the optimised variational distribution as discussed in \S~\ref{sub:4_1}. Here we have used the relation 
\begin{equation}
   p(z_i | \hat{z}_{p, i}) \propto p(\hat{z}_{p, i} | z_i) \int p( z_i | \boldsymbol{\phi}) p(\boldsymbol{\phi}) \, d\boldsymbol{\phi}_{\rm nz}\, ,
   \label{eq:posterior_indiv}
\end{equation}
where we omitted the variable $\sigma$ for simplicity.
Since many contemporary cluster mass estimation pipelines do not support probabilistic catalogs, we recommend to classify galaxies into the included and excluded bins using a cutoff at $\omega_i$ (here we choose 0.5). 
While using a catalog appropriately weighted by the bin-membership probability would be preferable, this approach is a good approximation if the posterior Eq.~\ref{eq:posterior_indiv} is not too broad or asymmetric which likely will be the case for upcoming surveys. 
If we run the unfolding algorithm again on the galaxies selected in this way, we obtain Fig.~\ref{fig:redshift_selection_removal}. (`VI [1, 99]\%', `Median VI') denote the corresponding percentile intervals and the median obtained by rerunning the deconvolution algorithm directly on the selected galaxies. The dashed red (median) and red points ([1, 99]\% percentile interval) denote the corresponding result from Eq.~\ref{eq:included_excluded_phi}. 
We can see that they are very close and using a catalog of galaxies classified according to the posterior redshift is appropriate. This also highlights our choice of the Dirichlet for our variational distribution because its properties allow analytical evaluation of Eq.~\ref{eq:included_excluded_phi} which is much faster than re-running a deconvolution algorithm. The sharp selection is a result of applying a deconvolution and will also be realized in the proposed simple application where we generate a catalog by classifying bin-membership based on Eq.~\ref{eq:posterior_indiv}.
In conclusion any subsequent analysis can then use said catalog in conjunction with the posterior sample redshift distribution. Since many contemporary algorithms do not support the usage of the full sample redshift distribution posterior, draws from the latter can be mapped to any parametrization the practitioner has implemented. Examples include marginalizing over uncertainty in the mean of the source sample redshift distribution. This can be achieved by sampling from the Dirichlet and calculating the mean of each of these draws, yielding a distribution over sample redshift distribution means. 

\begin{figure}
    \centering
    \includegraphics[scale=0.5]{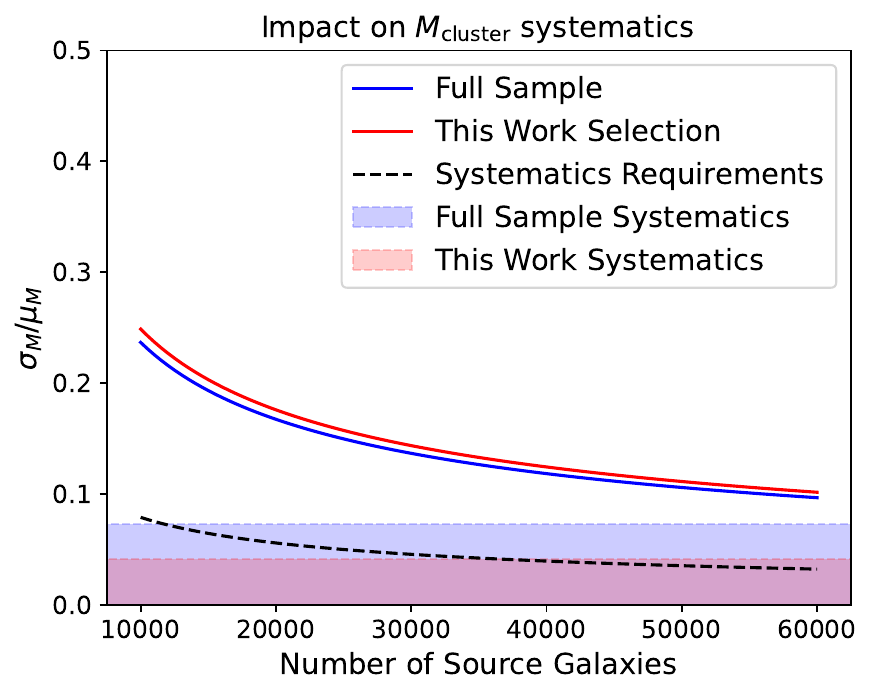}
    \caption{ Coefficient of variation for cluster mass $M$ as a function of the number of source galaxies. Statistical error: optimized selection (`VI [1, 99]\%' in Fig.~\ref{fig:redshift_selection_removal}) shown as the red line ('This Work Selection'), and the coefficient of variation of the full sample ('Full Sample'). The effective number of galaxies for the case `This Work Selection' will be 20\% lower than the corresponding value quoted on the x-axis. Systematic error requirements: the black dashed line ('Systematics Requirements') shows 1/3 of the statistical error of the `Full Sample'. Systematic error realized by the `Full Sample' and `This Work Selection' scenarios are illustrated by the horizontal areas.  }
    \label{fig:systematics_clusters}
\end{figure}

In the following we will denote:
\begin{equation}
    M \propto  \Sigma_{\rm crit} \left\langle \gamma_{T} \right \rangle
\end{equation}
where the estimate for the cluster mass $M$ is proportional to the product of the critical surface density $\Sigma_{\rm crit}$ and the tangential shear $\left\langle \gamma_{T} \right \rangle$ averaged over an angular bin and redshift. Both of these quantities are random variables, due to the uncertainty of the inferred sample redshift distribution. However for the large sample sizes in LSST the variance of $\Sigma_{\rm crit}$ can be expected to be quite small and will be neglected here \footnote{If this is not the case standard results for the product of two random variables are readily available in the statistics literature or can be estimated using bootstrap techniques.}. Fig.~\ref{fig:systematics_clusters} illustrates how the coefficient of variation for an estimate of $M$ changes as a function of the number of source galaxies. We show the coefficient of variation for optimized selection presented (`VI [1, 99]\%') as the red line (`This Work Selection') and the coefficient of variation of the full sample ('Full Sample') as the blue line. These lines constitute the statistical error budget. Note that we compare here scenarios before and after the removal of 20\% of the galaxy sample, so the effective number of galaxies for the case `This Work Selection' will be 20\% lower than the corresponding value quoted on the x-axis. It can be seen that due to a loss of 20\% of galaxies the statistical error on a cluster mass measurement is slightly higher. This scales approximately with $\sqrt{N}$, where $N$ is the sample size. The black dashed line (`Systematics Requirements') shows 1/3 of the statistical error of the `Full Sample' scenario for illustrative purposes. Ideally we would like to ensure that our systematic error budget, which in this analysis is dominated by photometric redshift systematics in $\Sigma_{\rm crit}$, are below this value. The systematic error budget from the two scenarios `Full Sample' and 'This Work Selection' are illustrated by the horizontal areas. Note that the systematic error will not decrease as a function of the number of source galaxies if it originated from a model misspecification error \footnote{Note that this is a conservative presentation since the unfolding estimate will improve with increasing number of galaxies, so we can expect that `This Work Selection' will improve with increasing number of galaxies in the sample.}. We can see that we can exploit a factor of 4 larger source galaxy samples, while still meeting our requirements on systematic error, if we optimize the sample selection to exclude redshift bins that are subject to model misspecification. 

This consideration illustrates that reducing model misspecification error (or the risk thereof) at a modest reduction in sample size is advisable in terms of improving the robustness of cluster mass measurements. Similar considerations also recently led to increased attention to optimized sample selection to robustify intrinsic alignment measurements \citep[see e.g.][]{2024arXiv241022272M} and where an integral part of the HSC Year 3 analysis \citep{2023MNRAS.524.5109R}. 

We particularly recommend our procedure in inference cases that lack model identifiability and scenarios with an incomplete model set, i.e. cases where the `true' model is not an element of the models being considered. As discussed this is a relevant scenario in photometric redshift estimation. We further require that while not being able to accurately quantify the exact extent of the model misspecification, we are able to localize it in the data and judge its potential impact using the described comparison of different modelling scenarios as a lower bound. While this is a significant, albeit unavoidable assumption, in practice it is often a fair one. 
Physical intuition and sensitivity studies often allow the localization of limitations in the model set. For example, photometric redshift outliers arise due to the limited capability to distinguish between known atomic lines in optical broad band photometry. As such any sample optimization needs to be tuned in light of physical insight and our methodology is no exception. 

\section{Summary and Conclusions}
\label{sec:6}
The mitigation of systematic errors is an important challenge in modern observational cosmology and many areas of fundamental science. The unprecedented size of cosmological datasets that will be observed by \textit{Euclid} and the Vera Rubin Observatory implies that the accuracy in the modelling of cosmological and astrophysical effects needs to improve at a similar rate. As is the case in many areas of fundamental science, the inference of parameters of interest can be challenging due to degeneracies induced by the survey design. In large area optical surveys, the limited information available from broad-band photometry can lead to large posterior variances for quantities of interest. Therefore, the parametrization of systematics using complex, many parameter models can lead to a very high-dimensional latent space. This can render inferences computationally expensive and sensitive to prior choices. 

This work proposes an alternative to this approach using weak lensing cluster mass estimation as an example. We use the fact that the error in the modelling of $\Sigma_\crit$ is a strong function of redshift and galaxy type and that abating systematic biases for a moderate reduction in sample size is attractive for these analyses. We ``remove'' histogram bins from the latent space to reduce the photometric redshift error in cluster mass estimates by constructing a mixture model of included/excluded galaxy populations. By optimizing the composition of this mixture model using the cross-entropy method, we are able to reduce the modelling bias in $\Sigma_\crit$ by 60-70\%, while maintaining 90\% of the source galaxy sample. Of course these numbers will strongly depend on the analysis in question. However, especially in scenarios where modelling error is a strong function of the latent parameters, this approach provides a
powerful opportunity to optimise the statistical model. 

We note that our method is different from imposing cuts on observed quantities like colours, as we still perform inference on the full data and afterwards optimise our selection in latent space. 
In future work we want to facilitate the selection in bins of redshift, tangential shear and other variables like cluster membership probability and include computational cost as a component to the loss function.

This will allow us to work towards a complete description of different sources of systematic bias and avoid spurious over- and underestimation of the $\Sigma_\crit$ integrand at different redshift intervals. As discussed in \S~\ref{sec:3}, this under- and overestimation can compensate and lead to a low estimated modelling error despite large residual systematics. Similar issues can arise upon including additional variables like \eg\ galaxy type into the model. Galaxy type impacts both the model error in (photometry-based) redshift inference as well as galaxy-dark matter bias modelling error. As a result ``removing'' a given redshift bin might cause a spurious reduction in the $\Sigma_\crit$ bias, if the cross-correlation calibration was also biased for this galaxy population. We finally note that the issues discussed here arise in multiple areas of science and as such the presented method will have applications beyond cluster cosmology, including in areas such as cosmological 2 point correlation function measurements.

In the advent of large-area photometric surveys like LSST and Euclid, the control of systematics will be vital. Optimizing the sample selection to the science case is critical to control these biasing effects and to facilitate cosmological analyses with well-controlled fidelity.

% Future surveys like Rubin, Roman and Euclid will revolutionize the field by observing galaxy clusters to unprecedented accuracy. The most important prerequisite to use these samples to test the cosmological standard model is the estimation of accurate cluster masses. Photometric redshift errors are currently a dominant systematic in these studies.  At the same time, the signal-to-noise in weak Lensing cluster mass measurements is typically quite high. Thus, abating systematic bias at moderate reduction in signal-to-noise is attractive. This is especially true considering the difficulties in calibrating photometric redshift error accurately for faint, high redshift samples that typically comprise the bulk of the source sample. 

% We devise a scheme to optimise the source sample selection in latent space and demonstrate in a mock study that our approach leads to a reduction in the systematic error in the critical surface mass density $\Sigma_{\rm crit}$ on the 60-70\% level, while maintaining 90\% of the galaxy source sample. 

\section*{Acknowledgements}
Work at Argonne National Laboratory was supported by the U.S. Department of Energy, Office of High Energy Physics. Argonne, a U.S. Department of Energy Office of Science Laboratory, is operated by UChicago Argonne LLC under contract no. DE-AC02-06CH11357. MMR and NR acknowledges the Laboratory Directed Research and Development (LDRD) funding from Argonne National Laboratory, provided by the Director, Office of Science, of the U.S. Department of Energy under Contract No. DE-AC02-06CH11357. Work at Argonne National Laboratory was also supported
under the U.S. Department of Energy contract DE-AC02-06CH11357.
%%%%%%%%%%%%%%%%%%%%%%%%%%%%%%%%%%%%%%%%%%%%%%%%%%
\section*{Data Availability}
The data and analysis products, as well as the software, will be made publicly available after paper acceptance or earlier upon reasonable request. 



\bibliographystyle{mnras}
\bibliography{example} % if your bibtex file is called example.bib

\begin{thebibliography}{}
\makeatletter
\relax
\def\mn@urlcharsother{\let\do\@makeother \do\$\do\&\do\#\do\^\do\_\do\%\do\~}
\def\mn@doi{\begingroup\mn@urlcharsother \@ifnextchar [ {\mn@doi@} {\mn@doi@[]}}
\def\mn@doi@[#1]#2{\def\@tempa{#1}\ifx\@tempa\@empty \href {http://dx.doi.org/#2} {doi:#2}\else \href {http://dx.doi.org/#2} {#1}\fi \endgroup}
\def\mn@eprint#1#2{\mn@eprint@#1:#2::\@nil}
\def\mn@eprint@arXiv#1{\href {http://arxiv.org/abs/#1} {{\tt arXiv:#1}}}
\def\mn@eprint@dblp#1{\href {http://dblp.uni-trier.de/rec/bibtex/#1.xml} {dblp:#1}}
\def\mn@eprint@#1:#2:#3:#4\@nil{\def\@tempa {#1}\def\@tempb {#2}\def\@tempc {#3}\ifx \@tempc \@empty \let \@tempc \@tempb \let \@tempb \@tempa \fi \ifx \@tempb \@empty \def\@tempb {arXiv}\fi \@ifundefined {mn@eprint@\@tempb}{\@tempb:\@tempc}{\expandafter \expandafter \csname mn@eprint@\@tempb\endcsname \expandafter{\@tempc}}}

\bibitem[\protect\citeauthoryear{{Abazajian} et~al.,}{{Abazajian} et~al.}{2016}]{2016arXiv161002743A}
{Abazajian} K.~N.,  et~al., 2016, \mn@doi [arXiv e-prints] {10.48550/arXiv.1610.02743}, \href {https://ui.adsabs.harvard.edu/abs/2016arXiv161002743A} {p. arXiv:1610.02743}

\bibitem[\protect\citeauthoryear{{Abbott} et~al.,}{{Abbott} et~al.}{2020}]{desy1_clcosmo}
{Abbott} T.~M.~C.,  et~al., 2020, \mn@doi [\prd] {10.1103/PhysRevD.102.023509}, \href {https://ui.adsabs.harvard.edu/abs/2020PhRvD.102b3509A} {102, 023509}

\bibitem[\protect\citeauthoryear{{Ade} et~al.,}{{Ade} et~al.}{2019}]{2019JCAP...02..056A}
{Ade} P.,  et~al., 2019, \mn@doi [\jcap] {10.1088/1475-7516/2019/02/056}, \href {https://ui.adsabs.harvard.edu/abs/2019JCAP...02..056A} {2019, 056}

\bibitem[\protect\citeauthoryear{{Aguena} et~al.,}{{Aguena} et~al.}{2021}]{2021MNRAS.508.6092A}
{Aguena} M.,  et~al., 2021, \mn@doi [\mnras] {10.1093/mnras/stab2764}, \href {https://ui.adsabs.harvard.edu/abs/2021MNRAS.508.6092A} {508, 6092}

\bibitem[\protect\citeauthoryear{{Alarcon} et~al.,}{{Alarcon} et~al.}{2020a}]{2020arXiv200711132A}
{Alarcon} A.,  et~al., 2020a, arXiv e-prints, \href {https://ui.adsabs.harvard.edu/abs/2020arXiv200711132A} {p. arXiv:2007.11132}

\bibitem[\protect\citeauthoryear{{Alarcon}, {S{\'a}nchez}, {Bernstein}  \& {Gazta{\~n}aga}}{{Alarcon} et~al.}{2020b}]{2019arXiv191007127A}
{Alarcon} A.,  {S{\'a}nchez} C.,  {Bernstein} G.~M.,   {Gazta{\~n}aga} E.,  2020b, \mn@doi [\mnras] {10.1093/mnras/staa2478}, \href {https://ui.adsabs.harvard.edu/abs/2020MNRAS.498.2614A} {498, 2614}

\bibitem[\protect\citeauthoryear{{Albrecht} et~al.,}{{Albrecht} et~al.}{2006}]{detf06}
{Albrecht} A.,  et~al., 2006, \mn@doi [arXiv e-prints] {10.48550/arXiv.astro-ph/0609591}, \href {https://ui.adsabs.harvard.edu/abs/2006astro.ph..9591A} {pp astro--ph/0609591}

\bibitem[\protect\citeauthoryear{{Allen}, {Evrard}  \& {Mantz}}{{Allen} et~al.}{2011}]{2011ARA&A..49..409A}
{Allen} S.~W.,  {Evrard} A.~E.,   {Mantz} A.~B.,  2011, \mn@doi [\araa] {10.1146/annurev-astro-081710-102514}, \href {https://ui.adsabs.harvard.edu/abs/2011ARA&A..49..409A} {49, 409}

\bibitem[\protect\citeauthoryear{{Applegate} et~al.,}{{Applegate} et~al.}{2014}]{2014MNRAS.439...48A}
{Applegate} D.~E.,  et~al., 2014, \mn@doi [\mnras] {10.1093/mnras/stt2129}, \href {https://ui.adsabs.harvard.edu/abs/2014MNRAS.439...48A} {439, 48}

\bibitem[\protect\citeauthoryear{{Arnouts}, {Cristiani}, {Moscardini}, {Matarrese}, {Lucchin}, {Fontana}  \& {Giallongo}}{{Arnouts} et~al.}{1999}]{1999MNRAS.310..540A}
{Arnouts} S.,  {Cristiani} S.,  {Moscardini} L.,  {Matarrese} S.,  {Lucchin} F.,  {Fontana} A.,   {Giallongo} E.,  1999, \mn@doi [\mnras] {10.1046/j.1365-8711.1999.02978.x}, \href {https://ui.adsabs.harvard.edu/\#abs/1999MNRAS.310..540A} {310, 540}

\bibitem[\protect\citeauthoryear{{Bartelmann} \& {Schneider}}{{Bartelmann} \& {Schneider}}{2001}]{2001PhR...340..291B}
{Bartelmann} M.,  {Schneider} P.,  2001, \mn@doi [\physrep] {10.1016/S0370-1573(00)00082-X}, \href {https://ui.adsabs.harvard.edu/abs/2001PhR...340..291B} {340, 291}

\bibitem[\protect\citeauthoryear{{Bellagamba}, {Roncarelli}, {Maturi}  \& {Moscardini}}{{Bellagamba} et~al.}{2018}]{amico}
{Bellagamba} F.,  {Roncarelli} M.,  {Maturi} M.,   {Moscardini} L.,  2018, \mn@doi [\mnras] {10.1093/mnras/stx2701}, \href {https://ui.adsabs.harvard.edu/abs/2018MNRAS.473.5221B} {473, 5221}

\bibitem[\protect\citeauthoryear{{Ben{\'\i}tez}}{{Ben{\'\i}tez}}{2000}]{2000ApJ...536..571B}
{Ben{\'\i}tez} N.,  2000, \mn@doi [\apj] {10.1086/308947}, \href {https://ui.adsabs.harvard.edu/\#abs/2000ApJ...536..571B} {536, 571}

\bibitem[\protect\citeauthoryear{{Benson} et~al.,}{{Benson} et~al.}{2014}]{2014SPIE.9153E..1PB}
{Benson} B.~A.,  et~al., 2014, in {Holland} W.~S.,  {Zmuidzinas} J.,  eds,  Society of Photo-Optical Instrumentation Engineers (SPIE) Conference Series Vol. 9153, Millimeter, Submillimeter, and Far-Infrared Detectors and Instrumentation for Astronomy VII. p. 91531P (\mn@eprint {arXiv} {1407.2973}), \mn@doi{10.1117/12.2057305}

\bibitem[\protect\citeauthoryear{{Bleem} et~al.,}{{Bleem} et~al.}{2024}]{bleem24}
{Bleem} L.~E.,  et~al., 2024, \mn@doi [The Open Journal of Astrophysics] {10.21105/astro.2311.07512}, \href {https://ui.adsabs.harvard.edu/abs/2024OJAp....7E..13B} {7, 13}

\bibitem[\protect\citeauthoryear{{Bocquet} et~al.,}{{Bocquet} et~al.}{2023}]{bocquet23i}
{Bocquet} S.,  et~al., 2023, \mn@doi [arXiv e-prints] {10.48550/arXiv.2310.12213}, \href {https://ui.adsabs.harvard.edu/abs/2023arXiv231012213B} {p. arXiv:2310.12213}

\bibitem[\protect\citeauthoryear{{Bocquet} et~al.,}{{Bocquet} et~al.}{2024}]{bocquet24ii}
{Bocquet} S.,  et~al., 2024, \mn@doi [arXiv e-prints] {10.48550/arXiv.2401.02075}, \href {https://ui.adsabs.harvard.edu/abs/2024arXiv240102075B} {p. arXiv:2401.02075}

\bibitem[\protect\citeauthoryear{{B{\"o}hringer} \& {Schartel}}{{B{\"o}hringer} \& {Schartel}}{2013}]{2013AN....334..482B}
{B{\"o}hringer} H.,  {Schartel} N.,  2013, \mn@doi [Astronomische Nachrichten] {10.1002/asna.201211886}, \href {https://ui.adsabs.harvard.edu/abs/2013AN....334..482B} {334, 482}

\bibitem[\protect\citeauthoryear{{Bonnett}}{{Bonnett}}{2015}]{2015MNRAS.449.1043B}
{Bonnett} C.,  2015, \mn@doi [\mnras] {10.1093/mnras/stv230}, \href {https://ui.adsabs.harvard.edu/\#abs/2015MNRAS.449.1043B} {449, 1043}

\bibitem[\protect\citeauthoryear{Botev, Kroese, Rubinstein  \& L’Ecuyer}{Botev et~al.}{2013}]{BOTEV201335}
Botev Z.~I.,  Kroese D.~P.,  Rubinstein R.~Y.,   L’Ecuyer P.,  2013, in Rao C.,  Govindaraju V.,  eds, Handbook of Statistics, Vol.~31, Handbook of Statistics.
Elsevier, pp 35--59, \mn@doi{https://doi.org/10.1016/B978-0-444-53859-8.00003-5}, \url {https://www.sciencedirect.com/science/article/pii/B9780444538598000035}

\bibitem[\protect\citeauthoryear{Buchs et~al.,}{Buchs et~al.}{2019a}]{10.1093/mnras/stz2162}
Buchs R.,  et~al., 2019a, \mn@doi [Monthly Notices of the Royal Astronomical Society] {10.1093/mnras/stz2162}, 489, 820

\bibitem[\protect\citeauthoryear{{Buchs} et~al.,}{{Buchs} et~al.}{2019b}]{2019MNRAS.489..820B}
{Buchs} R.,  et~al., 2019b, \mn@doi [\mnras] {10.1093/mnras/stz2162}, \href {https://ui.adsabs.harvard.edu/abs/2019MNRAS.489..820B} {489, 820}

\bibitem[\protect\citeauthoryear{{Bulbul} et~al.,}{{Bulbul} et~al.}{2024}]{bulbul24}
{Bulbul} E.,  et~al., 2024, \mn@doi [arXiv e-prints] {10.48550/arXiv.2402.08452}, \href {https://ui.adsabs.harvard.edu/abs/2024arXiv240208452B} {p. arXiv:2402.08452}

\bibitem[\protect\citeauthoryear{{Carrasco Kind} \& {Brunner}}{{Carrasco Kind} \& {Brunner}}{2013}]{2013MNRAS.432.1483C}
{Carrasco Kind} M.,  {Brunner} R.~J.,  2013, \mn@doi [\mnras] {10.1093/mnras/stt574}, \href {https://ui.adsabs.harvard.edu/abs/2013MNRAS.432.1483C} {432, 1483}

\bibitem[\protect\citeauthoryear{{Cataneo} \& {Rapetti}}{{Cataneo} \& {Rapetti}}{2018}]{2018IJMPD..2748006C}
{Cataneo} M.,  {Rapetti} D.,  2018, \mn@doi [International Journal of Modern Physics D] {10.1142/S0218271818480061}, \href {https://ui.adsabs.harvard.edu/abs/2018IJMPD..2748006C} {27, 1848006}

\bibitem[\protect\citeauthoryear{{Cawthon} et~al.,}{{Cawthon} et~al.}{2022}]{2022MNRAS.513.5517C}
{Cawthon} R.,  et~al., 2022, \mn@doi [\mnras] {10.1093/mnras/stac1160}, \href {https://ui.adsabs.harvard.edu/abs/2022MNRAS.513.5517C} {513, 5517}

\bibitem[\protect\citeauthoryear{{Clerc} et~al.,}{{Clerc} et~al.}{2024}]{clerc24}
{Clerc} N.,  et~al., 2024, \mn@doi [arXiv e-prints] {10.48550/arXiv.2402.08457}, \href {https://ui.adsabs.harvard.edu/abs/2024arXiv240208457C} {p. arXiv:2402.08457}

\bibitem[\protect\citeauthoryear{{Collister} \& {Lahav}}{{Collister} \& {Lahav}}{2004}]{2004PASP..116..345C}
{Collister} A.~A.,  {Lahav} O.,  2004, \mn@doi [Publications of the Astronomical Society of the Pacific] {10.1086/383254}, \href {https://ui.adsabs.harvard.edu/\#abs/2004PASP..116..345C} {116, 345}

\bibitem[\protect\citeauthoryear{{Conroy}}{{Conroy}}{2013}]{conroy13}
{Conroy} C.,  2013, \mn@doi [\araa] {10.1146/annurev-astro-082812-141017}, \href {https://ui.adsabs.harvard.edu/abs/2013ARA&A..51..393C} {51, 393}

\bibitem[\protect\citeauthoryear{{Costanzi} et~al.,}{{Costanzi} et~al.}{2019}]{costanzi19}
{Costanzi} M.,  et~al., 2019, \mn@doi [\mnras] {10.1093/mnras/sty2665}, \href {https://ui.adsabs.harvard.edu/abs/2019MNRAS.482..490C} {482, 490}

\bibitem[\protect\citeauthoryear{{Davis} et~al.,}{{Davis} et~al.}{2017}]{2017arXiv171002517D}
{Davis} C.,  et~al., 2017, arXiv e-prints, \href {https://ui.adsabs.harvard.edu/\#abs/2017arXiv171002517D} {p. arXiv:1710.02517}

\bibitem[\protect\citeauthoryear{{Euclid Collaboration} et~al.,}{{Euclid Collaboration} et~al.}{2019}]{2019A&A...627A..23E}
{Euclid Collaboration} et~al., 2019, \mn@doi [\aap] {10.1051/0004-6361/201935088}, \href {https://ui.adsabs.harvard.edu/abs/2019A&A...627A..23E} {627, A23}

\bibitem[\protect\citeauthoryear{{Feldmann} et~al.,}{{Feldmann} et~al.}{2006}]{2006MNRAS.372..565F}
{Feldmann} R.,  et~al., 2006, \mn@doi [\mnras] {10.1111/j.1365-2966.2006.10930.x}, \href {https://ui.adsabs.harvard.edu/\#abs/2006MNRAS.372..565F} {372, 565}

\bibitem[\protect\citeauthoryear{{Gatti} et~al.,}{{Gatti} et~al.}{2018}]{2018MNRAS.477.1664G}
{Gatti} M.,  et~al., 2018, \mn@doi [\mnras] {10.1093/mnras/sty466}, \href {https://ui.adsabs.harvard.edu/\#abs/2018MNRAS.477.1664G} {477, 1664}

\bibitem[\protect\citeauthoryear{{Gatti} et~al.,}{{Gatti} et~al.}{2022}]{2022MNRAS.510.1223G}
{Gatti} M.,  et~al., 2022, \mn@doi [\mnras] {10.1093/mnras/stab3311}, \href {https://ui.adsabs.harvard.edu/abs/2022MNRAS.510.1223G} {510, 1223}

\bibitem[\protect\citeauthoryear{Geiger \& Heckerman}{Geiger \& Heckerman}{1970}]{Geiger:1970}
Geiger D.,  Heckerman D.,  1970, \mn@doi [Annals of Statistics] {10.1214/aos/1069362752}, 25

\bibitem[\protect\citeauthoryear{{Gerdes}, {Sypniewski}, {McKay}, {Hao}, {Weis}, {Wechsler}  \& {Busha}}{{Gerdes} et~al.}{2010}]{2010ApJ...715..823G}
{Gerdes} D.~W.,  {Sypniewski} A.~J.,  {McKay} T.~A.,  {Hao} J.,  {Weis} M.~R.,  {Wechsler} R.~H.,   {Busha} M.~T.,  2010, \mn@doi [\apj] {10.1088/0004-637X/715/2/823}, \href {https://ui.adsabs.harvard.edu/\#abs/2010ApJ...715..823G} {715, 823}

\bibitem[\protect\citeauthoryear{{Ghirardini} et~al.,}{{Ghirardini} et~al.}{2024}]{Ghirardini24}
{Ghirardini} V.,  et~al., 2024, \mn@doi [arXiv e-prints] {10.48550/arXiv.2402.08458}, \href {https://ui.adsabs.harvard.edu/abs/2024arXiv240208458G} {p. arXiv:2402.08458}

\bibitem[\protect\citeauthoryear{{Giannini} et~al.,}{{Giannini} et~al.}{2024}]{2024MNRAS.527.2010G}
{Giannini} G.,  et~al., 2024, \mn@doi [\mnras] {10.1093/mnras/stad2945}, \href {https://ui.adsabs.harvard.edu/abs/2024MNRAS.527.2010G} {527, 2010}

\bibitem[\protect\citeauthoryear{{Graham}, {Connolly}, {Ivezi{\'c}}, {Schmidt}, {Jones}, {Juri{\'c}}, {Daniel}  \& {Yoachim}}{{Graham} et~al.}{2018}]{2018AJ....155....1G}
{Graham} M.~L.,  {Connolly} A.~J.,  {Ivezi{\'c}} {\v{Z}}.,  {Schmidt} S.~J.,  {Jones} R.~L.,  {Juri{\'c}} M.,  {Daniel} S.~F.,   {Yoachim} P.,  2018, \mn@doi [\aj] {10.3847/1538-3881/aa99d4}, \href {https://ui.adsabs.harvard.edu/abs/2018AJ....155....1G} {155, 1}

\bibitem[\protect\citeauthoryear{{Graham} et~al.,}{{Graham} et~al.}{2020}]{2020AJ....159..258G}
{Graham} M.~L.,  et~al., 2020, \mn@doi [\aj] {10.3847/1538-3881/ab8a43}, \href {https://ui.adsabs.harvard.edu/abs/2020AJ....159..258G} {159, 258}

\bibitem[\protect\citeauthoryear{{Grandis}, {Bocquet}, {Mohr}, {Klein}  \& {Dolag}}{{Grandis} et~al.}{2021}]{grandis21}
{Grandis} S.,  {Bocquet} S.,  {Mohr} J.~J.,  {Klein} M.,   {Dolag} K.,  2021, \mn@doi [\mnras] {10.1093/mnras/stab2414}, \href {https://ui.adsabs.harvard.edu/abs/2021MNRAS.507.5671G} {507, 5671}

\bibitem[\protect\citeauthoryear{{Greisel}, {Seitz}, {Drory}, {Bender}, {Saglia}  \& {Snigula}}{{Greisel} et~al.}{2015}]{2015MNRAS.451.1848G}
{Greisel} N.,  {Seitz} S.,  {Drory} N.,  {Bender} R.,  {Saglia} R.~P.,   {Snigula} J.,  2015, \mn@doi [\mnras] {10.1093/mnras/stv1005}, \href {http://adsabs.harvard.edu/abs/2015MNRAS.451.1848G} {451, 1848}

\bibitem[\protect\citeauthoryear{{Hartley} et~al.,}{{Hartley} et~al.}{2020}]{2020MNRAS.496.4769H}
{Hartley} W.~G.,  et~al., 2020, \mn@doi [\mnras] {10.1093/mnras/staa1812}, \href {https://ui.adsabs.harvard.edu/abs/2020MNRAS.496.4769H} {496, 4769}

\bibitem[\protect\citeauthoryear{{Hildebrandt} et~al.,}{{Hildebrandt} et~al.}{2021}]{2021A&A...647A.124H}
{Hildebrandt} H.,  et~al., 2021, \mn@doi [\aap] {10.1051/0004-6361/202039018}, \href {https://ui.adsabs.harvard.edu/abs/2021A&A...647A.124H} {647, A124}

\bibitem[\protect\citeauthoryear{{Hilton} et~al.,}{{Hilton} et~al.}{2021}]{hilton21}
{Hilton} M.,  et~al., 2021, \mn@doi [\apjs] {10.3847/1538-4365/abd023}, \href {https://ui.adsabs.harvard.edu/abs/2021ApJS..253....3H} {253, 3}

\bibitem[\protect\citeauthoryear{{Hoyle}}{{Hoyle}}{2016}]{2016A&C....16...34H}
{Hoyle} B.,  2016, \mn@doi [Astronomy and Computing] {10.1016/j.ascom.2016.03.006}, \href {https://ui.adsabs.harvard.edu/\#abs/2016A&C....16...34H} {16, 34}

\bibitem[\protect\citeauthoryear{Huterer, Lin, Busha, Wechsler  \& Cunha}{Huterer et~al.}{2014}]{10.1093/mnras/stu1424}
Huterer D.,  Lin H.,  Busha M.~T.,  Wechsler R.~H.,   Cunha C.~E.,  2014, \mn@doi [\mnras] {10.1093/mnras/stu1424}, 444, 129

\bibitem[\protect\citeauthoryear{{Ilbert} et~al.,}{{Ilbert} et~al.}{2006}]{2006A&A...457..841I}
{Ilbert} O.,  et~al., 2006, \mn@doi [\aap] {10.1051/0004-6361:20065138}, \href {http://adsabs.harvard.edu/abs/2006A%26A...457..841I} {457, 841}

\bibitem[\protect\citeauthoryear{Kullback \& Leibler}{Kullback \& Leibler}{1951}]{kullback1951information}
Kullback S.,  Leibler R.~A.,  1951, The annals of mathematical statistics, 22, 79

\bibitem[\protect\citeauthoryear{{LSST Science Collaboration} et~al.,}{{LSST Science Collaboration} et~al.}{2009}]{2009arXiv0912.0201L}
{LSST Science Collaboration} et~al., 2009, arXiv e-prints, \href {https://ui.adsabs.harvard.edu/\#abs/2009arXiv0912.0201L} {p. arXiv:0912.0201}

\bibitem[\protect\citeauthoryear{{Leistedt}, {Mortlock}  \& {Peiris}}{{Leistedt} et~al.}{2016}]{2016MNRAS.460.4258L}
{Leistedt} B.,  {Mortlock} D.~J.,   {Peiris} H.~V.,  2016, \mn@doi [\mnras] {10.1093/mnras/stw1304}, \href {http://adsabs.harvard.edu/abs/2016MNRAS.460.4258L} {460, 4258}

\bibitem[\protect\citeauthoryear{{Lesci} et~al.,}{{Lesci} et~al.}{2022a}]{kidsdr3_clcosmo}
{Lesci} G.~F.,  et~al., 2022a, \mn@doi [\aap] {10.1051/0004-6361/202040194}, \href {https://ui.adsabs.harvard.edu/abs/2022A&A...659A..88L} {659, A88}

\bibitem[\protect\citeauthoryear{{Lesci} et~al.,}{{Lesci} et~al.}{2022b}]{kidsdr3_clcat}
{Lesci} G.~F.,  et~al., 2022b, \mn@doi [\aap] {10.1051/0004-6361/202040194}, \href {https://ui.adsabs.harvard.edu/abs/2022A&A...659A..88L} {659, A88}

\bibitem[\protect\citeauthoryear{{Malz} \& {Hogg}}{{Malz} \& {Hogg}}{2020}]{2020arXiv200712178M}
{Malz} A.~I.,  {Hogg} D.~W.,  2020, arXiv e-prints, \href {https://ui.adsabs.harvard.edu/abs/2020arXiv200712178M} {p. arXiv:2007.12178}

\bibitem[\protect\citeauthoryear{{Masters}, {Stern}, {Cohen}, {Capak}, {Rhodes}, {Castander}  \& {Paltani}}{{Masters} et~al.}{2017}]{2017ApJ...841..111M}
{Masters} D.~C.,  {Stern} D.~K.,  {Cohen} J.~G.,  {Capak} P.~L.,  {Rhodes} J.~D.,  {Castander} F.~J.,   {Paltani} S.,  2017, \mn@doi [\apj] {10.3847/1538-4357/aa6f08}, \href {https://ui.adsabs.harvard.edu/abs/2017ApJ...841..111M} {841, 111}

\bibitem[\protect\citeauthoryear{{Masters} et~al.,}{{Masters} et~al.}{2019}]{2019ApJ...877...81M}
{Masters} D.~C.,  et~al., 2019, \mn@doi [\apj] {10.3847/1538-4357/ab184d}, \href {https://ui.adsabs.harvard.edu/abs/2019ApJ...877...81M} {877, 81}

\bibitem[\protect\citeauthoryear{{McCullough} et~al.,}{{McCullough} et~al.}{2024}]{2024arXiv241022272M}
{McCullough} J.,  et~al., 2024, \mn@doi [arXiv e-prints] {10.48550/arXiv.2410.22272}, \href {https://ui.adsabs.harvard.edu/abs/2024arXiv241022272M} {p. arXiv:2410.22272}

\bibitem[\protect\citeauthoryear{{McQuinn} \& {White}}{{McQuinn} \& {White}}{2013}]{2013MNRAS.433.2857M}
{McQuinn} M.,  {White} M.,  2013, \mn@doi [\mnras] {10.1093/mnras/stt914}, \href {https://ui.adsabs.harvard.edu/\#abs/2013MNRAS.433.2857M} {433, 2857}

\bibitem[\protect\citeauthoryear{{M{\'e}nard}, {Scranton}, {Schmidt}, {Morrison}, {Jeong}, {Budavari}  \& {Rahman}}{{M{\'e}nard} et~al.}{2013}]{2013arXiv1303.4722M}
{M{\'e}nard} B.,  {Scranton} R.,  {Schmidt} S.,  {Morrison} C.,  {Jeong} D.,  {Budavari} T.,   {Rahman} M.,  2013, arXiv e-prints, \href {https://ui.adsabs.harvard.edu/\#abs/2013arXiv1303.4722M} {p. arXiv:1303.4722}

\bibitem[\protect\citeauthoryear{{Morrison}, {Hildebrandt}, {Schmidt}, {Baldry}, {Bilicki}, {Choi}, {Erben}  \& {Schneider}}{{Morrison} et~al.}{2017}]{Morrison2016}
{Morrison} C.~B.,  {Hildebrandt} H.,  {Schmidt} S.~J.,  {Baldry} I.~K.,  {Bilicki} M.,  {Choi} A.,  {Erben} T.,   {Schneider} P.,  2017, \mn@doi [\mnras] {10.1093/mnras/stx342}, \href {https://ui.adsabs.harvard.edu/abs/2017MNRAS.467.3576M} {467, 3576}

\bibitem[\protect\citeauthoryear{{Mroczkowski} et~al.,}{{Mroczkowski} et~al.}{2019}]{2019SSRv..215...17M}
{Mroczkowski} T.,  et~al., 2019, \mn@doi [\ssr] {10.1007/s11214-019-0581-2}, \href {https://ui.adsabs.harvard.edu/abs/2019SSRv..215...17M} {215, 17}

\bibitem[\protect\citeauthoryear{{Myles} et~al.,}{{Myles} et~al.}{2021}]{2021MNRAS.505.4249M}
{Myles} J.,  et~al., 2021, \mn@doi [\mnras] {10.1093/mnras/stab1515}, \href {https://ui.adsabs.harvard.edu/abs/2021MNRAS.505.4249M} {505, 4249}

\bibitem[\protect\citeauthoryear{{Newman}}{{Newman}}{2008}]{2008ApJ...684...88N}
{Newman} J.~A.,  2008, \mn@doi [\apj] {10.1086/589982}, \href {https://ui.adsabs.harvard.edu/\#abs/2008ApJ...684...88N} {684, 88}

\bibitem[\protect\citeauthoryear{{Newman} \& {Gruen}}{{Newman} \& {Gruen}}{2022}]{2022ARA&A..60..363N}
{Newman} J.~A.,  {Gruen} D.,  2022, \mn@doi [\araa] {10.1146/annurev-astro-032122-014611}, \href {https://ui.adsabs.harvard.edu/abs/2022ARA&A..60..363N} {60, 363}

\bibitem[\protect\citeauthoryear{{Newman} et~al.,}{{Newman} et~al.}{2015}]{2015APh....63...81N}
{Newman} J.~A.,  et~al., 2015, \mn@doi [Astroparticle Physics] {10.1016/j.astropartphys.2014.06.007}, \href {https://ui.adsabs.harvard.edu/abs/2015APh....63...81N} {63, 81}

\bibitem[\protect\citeauthoryear{{Oke}}{{Oke}}{1974}]{oke74}
{Oke} J.~B.,  1974, \mn@doi [\apjs] {10.1086/190287}, \href {https://ui.adsabs.harvard.edu/abs/1974ApJS...27...21O} {27, 21}

\bibitem[\protect\citeauthoryear{{Planck Collaboration} et~al.,}{{Planck Collaboration} et~al.}{2016}]{planck15_clcosmo}
{Planck Collaboration} et~al., 2016, \mn@doi [\aap] {10.1051/0004-6361/201525833}, \href {https://ui.adsabs.harvard.edu/abs/2016A&A...594A..24P} {594, A24}

\bibitem[\protect\citeauthoryear{{Planck Collaboration} et~al.,}{{Planck Collaboration} et~al.}{2020}]{2020A&A...641A...6P}
{Planck Collaboration} et~al., 2020, \mn@doi [\aap] {10.1051/0004-6361/201833910}, \href {https://ui.adsabs.harvard.edu/abs/2020A&A...641A...6P} {641, A6}

\bibitem[\protect\citeauthoryear{{Pratt}, {Arnaud}, {Biviano}, {Eckert}, {Ettori}, {Nagai}, {Okabe}  \& {Reiprich}}{{Pratt} et~al.}{2019}]{2019SSRv..215...25P}
{Pratt} G.~W.,  {Arnaud} M.,  {Biviano} A.,  {Eckert} D.,  {Ettori} S.,  {Nagai} D.,  {Okabe} N.,   {Reiprich} T.~H.,  2019, \mn@doi [\ssr] {10.1007/s11214-019-0591-0}, \href {https://ui.adsabs.harvard.edu/abs/2019SSRv..215...25P} {215, 25}

\bibitem[\protect\citeauthoryear{Raccanelli, Rahman  \& Kovetz}{Raccanelli et~al.}{2017}]{10.1093/mnras/stx691}
Raccanelli A.,  Rahman M.,   Kovetz E.~D.,  2017, \mn@doi [\mnras] {10.1093/mnras/stx691}, 468, 3650

\bibitem[\protect\citeauthoryear{{Rau}, {Seitz}, {Brimioulle}, {Frank}, {Friedrich}, {Gruen}  \& {Hoyle}}{{Rau} et~al.}{2015}]{2015MNRAS.452.3710R}
{Rau} M.~M.,  {Seitz} S.,  {Brimioulle} F.,  {Frank} E.,  {Friedrich} O.,  {Gruen} D.,   {Hoyle} B.,  2015, \mn@doi [\mnras] {10.1093/mnras/stv1567}, \href {https://ui.adsabs.harvard.edu/abs/2015MNRAS.452.3710R} {452, 3710}

\bibitem[\protect\citeauthoryear{{Rau}, {Wilson}  \& {Mandelbaum}}{{Rau} et~al.}{2020}]{2020MNRAS.491.4768R}
{Rau} M.~M.,  {Wilson} S.,   {Mandelbaum} R.,  2020, \mn@doi [\mnras] {10.1093/mnras/stz3295}, \href {https://ui.adsabs.harvard.edu/abs/2020MNRAS.491.4768R} {491, 4768}

\bibitem[\protect\citeauthoryear{{Rau}, {Morrison}, {Schmidt}, {Wilson}, {Mandelbaum}, {Mao}, {Mao}  \& {LSST Dark Energy Science Collaboration}}{{Rau} et~al.}{2022}]{2022MNRAS.509.4886R}
{Rau} M.~M.,  {Morrison} C.~B.,  {Schmidt} S.~J.,  {Wilson} S.,  {Mandelbaum} R.,  {Mao} Y.~Y.,  {Mao} Y.~Y.,   {LSST Dark Energy Science Collaboration} 2022, \mn@doi [\mnras] {10.1093/mnras/stab3290}, \href {https://ui.adsabs.harvard.edu/abs/2022MNRAS.509.4886R} {509, 4886}

\bibitem[\protect\citeauthoryear{{Rau} et~al.,}{{Rau} et~al.}{2023}]{2023MNRAS.524.5109R}
{Rau} M.~M.,  et~al., 2023, \mn@doi [\mnras] {10.1093/mnras/stad1962}, \href {https://ui.adsabs.harvard.edu/abs/2023MNRAS.524.5109R} {524, 5109}

\bibitem[\protect\citeauthoryear{Rubinstein}{Rubinstein}{1999}]{rubinstein1999}
Rubinstein R.,  1999, \mn@doi [Methodology And Computing In Applied Probability] {10.1023/A:1010091220143}, 1, 127

\bibitem[\protect\citeauthoryear{{Rykoff} et~al.,}{{Rykoff} et~al.}{2014}]{redmapper}
{Rykoff} E.~S.,  et~al., 2014, \mn@doi [\apj] {10.1088/0004-637X/785/2/104}, \href {https://ui.adsabs.harvard.edu/abs/2014ApJ...785..104R} {785, 104}

\bibitem[\protect\citeauthoryear{{Rykoff} et~al.,}{{Rykoff} et~al.}{2016}]{desy1clcat}
{Rykoff} E.~S.,  et~al., 2016, \mn@doi [\apjs] {10.3847/0067-0049/224/1/1}, \href {https://ui.adsabs.harvard.edu/abs/2016ApJS..224....1R} {224, 1}

\bibitem[\protect\citeauthoryear{{Salvati} et~al.,}{{Salvati} et~al.}{2022}]{salvati22}
{Salvati} L.,  et~al., 2022, \mn@doi [\apj] {10.3847/1538-4357/ac7ab4}, \href {https://ui.adsabs.harvard.edu/abs/2022ApJ...934..129S} {934, 129}

\bibitem[\protect\citeauthoryear{{Salvato}, {Ilbert}  \& {Hoyle}}{{Salvato} et~al.}{2019}]{2019NatAs...3..212S}
{Salvato} M.,  {Ilbert} O.,   {Hoyle} B.,  2019, \mn@doi [Nature Astronomy] {10.1038/s41550-018-0478-0}, \href {https://ui.adsabs.harvard.edu/abs/2019NatAs...3..212S} {3, 212}

\bibitem[\protect\citeauthoryear{{S{\'a}nchez} \& {Bernstein}}{{S{\'a}nchez} \& {Bernstein}}{2019}]{2019MNRAS.483.2801S}
{S{\'a}nchez} C.,  {Bernstein} G.~M.,  2019, \mn@doi [\mnras] {10.1093/mnras/sty3222}, \href {https://ui.adsabs.harvard.edu/\#abs/2019MNRAS.483.2801S} {483, 2801}

\bibitem[\protect\citeauthoryear{{Scottez} et~al.,}{{Scottez} et~al.}{2016}]{2016MNRAS.462.1683S}
{Scottez} V.,  et~al., 2016, \mn@doi [\mnras] {10.1093/mnras/stw1500}, \href {https://ui.adsabs.harvard.edu/\#abs/2016MNRAS.462.1683S} {462, 1683}

\bibitem[\protect\citeauthoryear{{Sommer}, {Schrabback}, {Applegate}, {Hilbert}, {Ansarinejad}, {Floyd}  \& {Grandis}}{{Sommer} et~al.}{2022}]{sommer22}
{Sommer} M.~W.,  {Schrabback} T.,  {Applegate} D.~E.,  {Hilbert} S.,  {Ansarinejad} B.,  {Floyd} B.,   {Grandis} S.,  2022, \mn@doi [\mnras] {10.1093/mnras/stab3052}, \href {https://ui.adsabs.harvard.edu/abs/2022MNRAS.509.1127S} {509, 1127}

\bibitem[\protect\citeauthoryear{{Sunyaev} \& {Zel'dovich}}{{Sunyaev} \& {Zel'dovich}}{1972}]{sunyaev72}
{Sunyaev} R.~A.,  {Zel'dovich} Y.~B.,  1972, Comments on Astrophysics and Space Physics, \href {http://adsabs.harvard.edu/cgi-bin/nph-bib_query?bibcode=1972CoASP...4..173S&amp;db_key=AST} {4, 173}

\bibitem[\protect\citeauthoryear{{Tagliaferri}, {Longo}, {Andreon}, {Capozziello}, {Donalek}  \& {Giordano}}{{Tagliaferri} et~al.}{2003}]{2003LNCS.2859..226T}
{Tagliaferri} R.,  {Longo} G.,  {Andreon} S.,  {Capozziello} S.,  {Donalek} C.,   {Giordano} G.,  2003, {Neural Networks for Photometric Redshifts Evaluation}.
pp 226--234, \mn@doi{10.1007/978-3-540-45216-4_26}

\bibitem[\protect\citeauthoryear{{The LSST Dark Energy Science Collaboration} et~al.,}{{The LSST Dark Energy Science Collaboration} et~al.}{2018}]{2018arXiv180901669T}
{The LSST Dark Energy Science Collaboration} et~al., 2018, \mn@doi [arXiv e-prints] {10.48550/arXiv.1809.01669}, \href {https://ui.adsabs.harvard.edu/abs/2018arXiv180901669T} {p. arXiv:1809.01669}

\bibitem[\protect\citeauthoryear{{Umetsu}}{{Umetsu}}{2020}]{umetsu20}
{Umetsu} K.,  2020, \mn@doi [\aapr] {10.1007/s00159-020-00129-w}, \href {https://ui.adsabs.harvard.edu/abs/2020A&ARv..28....7U} {28, 7}

\bibitem[\protect\citeauthoryear{{Voit}}{{Voit}}{2005}]{2005RvMP...77..207V}
{Voit} G.~M.,  2005, \mn@doi [Reviews of Modern Physics] {10.1103/RevModPhys.77.207}, \href {https://ui.adsabs.harvard.edu/abs/2005RvMP...77..207V} {77, 207}

\bibitem[\protect\citeauthoryear{White}{White}{1982}]{5805f73c-4dfa-385e-bd6d-68424fb9f5be}
White H.,  1982, Econometrica, 50, 1

\bibitem[\protect\citeauthoryear{{van den Busch} et~al.,}{{van den Busch} et~al.}{2020}]{2020A&A...642A.200V}
{van den Busch} J.~L.,  et~al., 2020, \mn@doi [\aap] {10.1051/0004-6361/202038835}, \href {https://ui.adsabs.harvard.edu/abs/2020A&A...642A.200V} {642, A200}

\makeatother
\end{thebibliography}

%\begin{thebibliography}{99}
%\end{thebibliography}

%%%%%%%%%%%%%%%%%%%%%%%%%%%%%%%%%%%%%%%%%%%%%%%%%%

%%%%%%%%%%%%%%%%% APPENDICES %%%%%%%%%%%%%%%%%%%%%

\appendix
\section{Mock Simulation}
\label{sec:mock_sim}
\subsection{Shape catalog}
We use the CLMM\footnote{\url{https://github.com/LSSTDESC/CLMM}} library \citep{2021MNRAS.508.6092A} to generate a mock realisation of the weak lensing observables of a cluster-scale halo.
We adopt a fiducial cosmology from the \textit{Planck} 2018 measurements of the cosmic microwave background \citep{2020A&A...641A...6P} and model a halo following a Navarro-Frenk-White (NFW) matter distribution with mass $M_{200c} = 5 \times 10^{14} \, M_\odot$ and concentration $c_{200c} = 5$ at redshift $z=0.5$.
We then use CLMM's \texttt{generate\_galaxy\_catalog()} method\footnote{\url{https://lsstdesc.org/CLMM/compiled-examples/demo_mock_cluster.html}} \citep[described in \S3.5.1 of][]{2021MNRAS.508.6092A} to generate a mock catalogue of $10^4$ background galaxies within a $(10 \, {\rm Mpc} \times 10 \, {\rm Mpc}$ square (at the cluster's redshift, which corresponds to a $(26.5' \times 26.5')$ patch of the sky).
The (true, spectroscopic) redshifts of the synthetic galaxies are drawn from the LSST/DESC Science Requirement Document galaxy redshift distribution \cite{2018arXiv180901669T}, enforcing $z \in [0.6, \, 3]$.
Photometric redshifts attribution is discussed in the next section (\S\ref{sec:app:sim:pz}).
Each galaxy is then assigned tangential and cross shear components, $g_+$ and $g_\times$, using a Gaussian shape noise with standard deviation $\sigma = 0.26$ \citep[expected for the LSST survey; see, \eg,][]{2018arXiv180901669T}.

\subsection{Photometric redshift catalog} \label{sec:app:sim:pz}
We generate three photometric redshift analysis products: 
\begin{itemize}[leftmargin=*]
    \item A fiducial photometric redshift estimate M1
    \item An alternative photometric redshift estimate M2
    \item An estimate for the sample redshift distribution based on spatial cross-correlations, WX
\end{itemize}
We consider 2 sources of photometric redshift error: additive statistical noise of $\sigma_z = 0.05$ and and a fraction of catastrophic outliers generated by superimposing a normal distribution with a mean of 2.5 and a standard deviation of 0.3.  

Catastrophic outliers arise in cases where two templates for the spectral energy distribution of galaxies fit to the measured photometry. Due to the position of optical filters and the position of atomic lines, this mismatch is expected to arise in LSST between redshifts of $z\approx 0.5$ and $z\approx 2.5$ \citep{2018AJ....155....1G}. In extreme cases like the scenario illustrated in Fig.1 in \citet{2019MNRAS.489..820B}, a maximum likelihood estimate of redshift is no longer unique, which is referred to as `lack of identifiability'. This uniqueness is a necessary assumption for consistency of the estimate \citep[see e.g.][]{5805f73c-4dfa-385e-bd6d-68424fb9f5be}. Since the data likelihood is unable to distinguish between solutions, inferences would rely on prior information to constrain redshifts, i.e. we are in a prior dominated scenario. Thus we often find competing models that produce different values for outlier rates. Following this argument we assume that our fiducial method (M1) is reporting a 10\% outlier fraction which is consistent with typical values \citep{2020AJ....159..258G} and the alternative method is reporting 5\% outliers. This is chosen to mimic the aforementioned differences in prior choices. We also consider a spatial cross-correlation method assumed here to be centred around the true photometric redshift. The errorbars ($\sigma_{\rm WX} = 0.003$) are selected in agreement with requirements on the calibration of the mean of the source sample redshift distribution formulated in \citet{2018arXiv180901669T}.

% Don't change these lines
\bsp	% typesetting comment
\label{lastpage}
\end{document}